\documentclass[dvips,graphicx,11pt]{article}
\usepackage{lscape,graphicx}
\usepackage{color}
\usepackage{amssymb}
\usepackage{euler}

\usepackage{mathptmx}      
\usepackage{natbib}  

\newcommand{\aap}{{Astron. Astrophys.}}
\newcommand{\apj}{{Astrophys. J.}}
\newcommand{\jgr}{{J. Geophys. Res.}}

\newcommand{\solphys}{{Solar Phys.}}

\usepackage{natbib}
\usepackage{pifont}
\usepackage{amsmath}

\begin{document}

\begin{center}
{\large\bf 

{The behaviour of magnetoacoustic waves in the neighbourhood of a two-dimensional null point: initially cylindrically-symmetric perturbations}}

\bigskip
{\large\bf {J.~A.~{McLaughlin} } }

\bigskip
{Department of Mathematics and Information Sciences, Northumbria University, Newcastle upon Tyne, NE1 8ST, United Kingdom}
\end{center}

\bigskip
{\bf Abstract:}

{

The propagation of magnetoacoustic waves in the neighbourhood of a 2D null point is investigated for both $\beta=0$ and $\beta \neq 0$ plasmas. Previous work has shown that the Alfv\'en speed, here $v_A \propto r$, plays a vital role in such systems and so a natural choice is to switch to polar coordinates. For the $\beta=0$ plasma, we derive an analytical solution for the behaviour of the fast magnetoacoustic wave in terms of the Klein-Gordon equation. We also solve the system with a semi-analytical WKB approximation which shows that the $\beta=0$ wave focuses on the null and contracts around it but, due to exponential decay, never  reaches the null  in a finite time.  For the $\beta \neq 0$ plasma, we solve the system numerically and find the behaviour to be similar to that of the $\beta=0$ system at large radii, but completely different close to the null. We show that for an initially cylindrically-symmetric fast magnetoacoustic wave perturbation, there is a decrease in wave speed along the separatrices and so the perturbation starts to take on a quasi-diamond shape; with the corners located along the separatrices. This is due to the growth in pressure gradients that reach a maximum along the separatrices, which in  turn  reduces the acceleration of the fast wave along the separatrices leading to a deformation of the wave morphology.

}




\section{Introduction}

MHD waves are ubiquitous in the Sun's atmosphere (e.g. Tomczyk {\emph{et al.}} \citeyear{Tomczyk2007}) and a variety of  observations have now  demonstrated the existence of wave activity for the three fundamental MHD wave modes: namely Alfv\'en waves and fast and slow magnetoacoustic waves.  Non-thermal line broadening and narrowing  due to  Alfv\'en waves has been reported by various authors, including Banerjee {\emph{et al.}} (\citeyear{Banerjee1998}), Erd\'elyi  {\emph{et al.}} (\citeyear{Erdelyi1998}),  Harrison {\emph{et al.}} (\citeyear{Harrison2002}) and O'Shea {\emph{et al.}} (\citeyear{OShea2003}; \citeyear{OShea2005}) and investigated both analytically (e.g. Dwivedi  \& Srivastava \citeyear{DS2006}) and numerically (e.g. Chmielewski  {\emph{et al.}} {\citeyear{Chmielewski2013}}, and references therein).

MHD wave behaviour is influenced strongly by the underlying magnetic structure (topology) and so it is useful to look at the topology itself. Potential field extrapolations of the coronal magnetic field can be made from photospheric magnetograms and such extrapolations show the existence of two key features of the magnetic topology: {\emph{magentic null points}} and  {\emph{separatrices}}. {{Null points}} are  weaknesses in the magnetic field at which  the field strength, and thus the Alfv\'en speed, is zero. {{Separatrices}} are topological features that separate regions of different magnetic connectivity and are an inevitable consequence of the isolated magnetic flux fragments  in the photosphere. Detailed investigations of the coronal magnetic field, using such potential field calculations, can be found in Beveridge     {\it{et al.}} (\citeyear{Beveridge2002}) and Brown \& Priest (\citeyear{Brown2001}). The number of resultant null points  depends upon the complexity of the magnetic flux distribution and tens of thousands are estimated to be present (see, e.g., Close {\emph{et al.}} \citeyear{Close2004}; Longcope \citeyear{L2005}; R{\'e}gnier {\emph{et al.}} \citeyear{RPH2008}; Longcope \& Parnell \citeyear{LP2009}).

MHD waves and magnetic topology {\emph{will}} encounter each other in the solar corona, e.g. waves emanating from a flare or CME will at some point encounter a coronal null point. MHD wave propagation within an inhomogeneous magnetic  medium  is a fundamental plasma process and the study of MHD wave behaviour in the neighbourhood of magnetic null points directly contributes to this area; see McLaughlin {\emph{et al.}} (\citeyear{McLaughlinREVIEW}) for a comprehensive review of the topic.

The behaviour of linear MHD waves, both magnetoacoustic waves and Alfv\'en waves, has been investigated in the neighbourhood of a variety of 2D null points (e.g. McLaughlin \& Hood \citeyear{MH2004}; \citeyear{MH2005}; \citeyear{MH2006a}; \citeyear{MH2006b}; McLaughlin \citeyear{M2013}). Nonlinear and 3D  MHD wave activity about coronal null points has also been investigated (e.g. Galsgaard {\emph{et al.}} \citeyear{Galsgaard2003};   Pontin \& {Galsgaard} \citeyear{PG2007};  Pontin {\emph{et al.}} \citeyear{PBG2007}; McLaughlin {\emph{et al.}} \citeyear{MFH2008};  \citeyear{McLaughlin2009}; Galsgaard \& Pontin \citeyear{klaus2011a}; \citeyear{klaus2011b}; Thurgood \& McLaughlin \citeyear{Thurgood2012}; \citeyear{Thurgood2013a}; \citeyear{Thurgood2013b}).

{{

Authors have also considered an X-point magnetic field configuration with a longitudinal (along the X-line) magnetic field $B_\parallel$. This has the effect that now the fast magnetoacoustic wave and Alfv\'en wave are linearly coupled by the gradients in the field. McClements {\it{et al.}} (\citeyear{McClements2006}) investigated such a coupling with a weak longitudinal guide field present ($B_\parallel \ll B_\perp$) and  Ben Ayed {\it{et al.}} (\citeyear{BenAyed2009}) considered a strong guide-field ($B_\parallel \gg B_\perp$). These authors found that the Alfv\'en wave is coupled into the fast mode, with the coupling strongest  on the separatrices and far from the X-line. In the limit of $B_\parallel \to 0$, the two modes are decoupled and the results of 2D work are recovered. More recently, {Ku{\'z}ma} {\it{et al.}} (\citeyear{KMS2015}) investigated similar coupling for a X-line formed  above two magnetic arcades, but now embedded in a model solar atmosphere with a realistic temperature distribution. They found that the formation of the Alfv\'en waves at the initial phase of temporal evolution is followed by  linear coupling between Alfv\'en and magnetoacoustic waves at a later time. The Alfv\'en waves  also  experience phase mixing and  scattering from inhomogeneous regions of Alfv\'en speed, and partial reflection from the model transition region.

}}

It is also clear that the plasma-$\beta$, i.e. the ratio of thermal plasma pressure to magnetic pressure, plays a key role. A very detailed and comprehensive set of 2D numerical simulations of wave propagation in a stratified magneto-atmosphere was conducted by Rosenthal {\it{et al.}} (\citeyear{Rosenthal2002}) and Bogdan   {\it{et al.}}       (\citeyear{Bogdan2003}). In these simulations, an oscillating piston generated both fast and slow MHD waves on a lower boundary and sent these waves up into the stratified magnetized plasma. Their calculations showed there was coupling between the fast and slow waves, and that this coupling was confined to a thin layer where the sound speed and the Alfv\'en velocity are comparable in magnitude, i.e. where the plasma-$\beta$ approaches unity. Away from this conversion zone, the waves were decoupled as either the magnetic pressure or thermal plasma pressure dominated. One of the aims of these papers was to see how the topology affected the propagation of waves, with the ratio of the sound speed to the Alfv\'en speed varying along every magnetic line of force. In this, their work and ours have the same ultimately goal; a fully 3D understanding of MHD wave propagation in the solar corona.

Other authors have also looked at MHD mode coupling: Cally \& Bogdan (\citeyear{Cally1997}) describes 2D simulations in which both $f$-modes and $p$-modes are partially converted to slow magnetoacoustic gravity waves due to strong gravitational stratification. De Moortel {\it{et al.}} (2004) investigated driving slow waves on the boundary of a 2D geometry with a horizontal density variation, where they found coupling between slow and fast waves and phase mixing of the slow waves. The coupling of different wave modes has also been investigated by  {Ferraro} \& {Plumpton} (\citeyear{Ferraro1958}), with Meijer G-functions by {Zhugzhd} \& {Dzhalilov} (\citeyear{Zhugzhd}), and with hypergeometric $_2F_3$ functions by Cally (\citeyear{Cally2001}). All these works considered mode coupling through a gravitational stratification, i.e. a vertical density inhomogenity. Finally, the coupling of fast waves and Alfv\'en waves has been investigated by Parker (\citeyear{Parker1991}) for linear MHD with a density gradient and by Nakariakov {\it{et al.}} (\citeyear{Valery1997}) for nonlinear excitation.

In this paper, we will investigate the behaviour of magnetoacoustic waves within inhomogeneous magnetic media. We will concentrate our investigations on wave behaviour excited via initially cylindrical-symmetric perturbations. Our paper has two aims: Firstly, we will investigate the behaviour of (fast) magnetoacoustic waves in a $\beta=0$ plasma using numerical, analytical and semi-analytical techniques. Secondly, we lift the $\beta=0$ assumption and study a $\beta \neq 0$ plasma. This naturally introduces slow waves to the system and so we will  investigate the behaviour of both types of  magnetoacoustic waves around a null point.



Two papers are key to our investigation: Firstly, McLaughlin \& Hood (\citeyear{MH2004}) investigated the behaviour of the fast magnetoacoustic wave in a $\beta=0$ plasma within a Cartesian geometry. They found that the fast magnetoacoustic wave was attracted to the null via a refraction effect and that all the wave energy accumulated at the null. Secondly,  McLaughlin \& Hood (\citeyear{MH2006b}) extended the 2004 model to include plasma pressure in  a $\beta \neq 0$ system. This led to the introduction of slow magnetoacoustic waves and coupling between the two types of  magnetoacoustic waves. However, the resultant behaviour was extremely complex and the investigate was again limited to a  Cartesian geometry. In this paper, we will investigate  the behaviour of magnetoacoustic waves in a $\beta \neq 0$ plasma  excited via initially cylindrical-symmetric perturbations. It is hoped that our  results will help begin to explain the complex resultant behaviour observed in  McLaughlin \& Hood (\citeyear{MH2006b}) and hence contribute to the overall  understanding of MHD mode conversion across the $\beta=1$ layer.

Our paper has the following outline:  The basic setup, equations and assumptions are described in $\S\ref{sec:1.1}$. The analytical, numerical and semi-analytical results for a $\beta=0$ plasma are presented in $\S\ref{sec:2.3.3}$ and the numerical results for a $\beta \neq 0$ plasma appear in $\S\ref{polarcoorindatesbetaneq0}$. The discussions and conclusions are given in $\S\ref{sec:6.10}$.


\section{Basic Equations}\label{sec:1.1}

We utilize the usual MHD equations appropriate to the solar corona, with pressure and resistivity included. Hence
\begin{eqnarray}
  \rho \left[ {\partial {\bf{v}}\over \partial t} + \left( {\bf{v}}\cdot\nabla \right) {\bf{v}} \right] &=& {  \frac{1}{\mu}}\left(\nabla \times {\bf{B}}\right)\times {\bf{B}}- \nabla p \; ,\nonumber \\
{\partial {\bf{B}}\over \partial t} &=& \nabla \times \left({\bf{v}}\times {\bf{B}}\right ) + \eta \nabla^2{\bf{B}} \; ,\nonumber \\
{\partial \rho\over \partial t} + \nabla \cdot \left (\rho {\bf{v}}\right) &=& 0 \; , \nonumber \\
{\partial p\over \partial t} + \left( \bf{v} \cdot \nabla \right) p &=& - \gamma p \nabla \cdot \bf{v}\; , \label{eq:1.1}
\end{eqnarray}
where $\rho$ is the mass density, ${\bf{v}}$ is the plasma velocity, ${\bf{B}}$ the magnetic induction (usually called the magnetic field), $p$ is the thermal plasma pressure, $ \mu = 4 \pi \times 10^{-7} \/\;\mathrm{Hm^{-1}}$ is  the magnetic permeability, $\eta = 1/\mu\sigma$ is the magnetic diffusivity in $ \mathrm{m}^2\mathrm{s}^{-1}$ and $\sigma$ the electrical conductivity. We have also neglected viscous terms in equations (\ref{eq:1.1}). Investigations involving viscous magnetofluids can be found in  Kumar \& Bhattacharyya  (\citeyear{KB2011}) and McLaughlin {\emph{et al.}} (\citeyear{McLaughlin2011b}) and references therein.


\subsection{Basic equilibrium}\label{sec:1.2}

\begin{figure}[t]
\begin{center}
\includegraphics[width=2.15in]{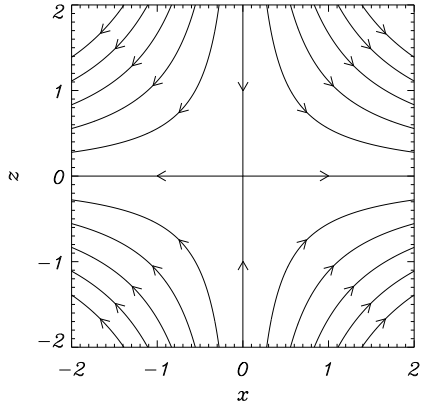}
\caption{Our choice of equilibrium magnetic field.}
\label{finitebetaXpoint}
\end{center}
\end{figure}

The basic magnetic field structure is taken as a simple 2D X-type null point. Therefore, the magnetic field is taken as
\begin{equation}
{\bf{B}}_0 = \frac{B}{L} \left(x, 0, -z\right),\label{eq:1.2}
\end{equation}
where $B$ is a characteristic field strength and $L$ is the length scale for magnetic field variations. This magnetic field can be seen in Figure \ref{finitebetaXpoint}. Obviously, this magnetic configuration is no longer valid far away from the null point since the field strength tends to infinity. However, McLaughlin \& Hood (\citeyear{MH2006a}) looked at a magnetic field that decays far from the null (for a $\beta=0$ plasma) and they found that the key results from McLaughlin \& Hood (\citeyear{MH2004}) remain true very close to the null. In addition, equation (\ref{eq:1.2}) is potential, although in general coronal fields are  twisted and thus a potential field is a coarse approximation. {{The aim of studying waves in a 2D configuration is one of simplicity: there are a lot of complicated effects including mode transition and coupling, and a 2D geometry allows us to better understand and explain these behaviours before the extension to 3D. Our modelling philosophy is to build up our models incrementally, with an emphasis on understanding the underlying physical processes at each step, since (as detailed in the introduction) the solar corona is extremely inhomogeneous in all its characteristics.}}


In this paper, the linearized MHD equations are used to study the nature of magnetoacoustic wave propagation near the null point. Using subscript $0$ for equilibrium quantities (e.g. ${\bf{B}}_0$), $\mathbf{b}$ to denote perturbed magnetic field and subscript $1$ for all other perturbed quantities, the linearized equation of motion becomes
\begin{equation}\label{eq:2.3}
 \rho_0 \frac{\partial \mathbf{v}_1}{\partial t} = \left(\frac{ \nabla \times \mathbf{b}}{\mu} \right) \times \mathbf{B}_0 - \nabla p_1\; ,
\end{equation}
the linearized induction equation
\begin{equation}\label{eq:2.4}
\qquad    {\partial {\bf{b}}\over \partial t} = \nabla \times
    ({\bf{v}}_1 \times {\bf{B}}_0)  + \eta \nabla^2 {{\bf{b}}}\; ,
\end{equation}
the linearized equation of mass continuity
\begin{equation}\label{eq:2.5}
\frac{\partial \rho_1} {\partial t} + \nabla\cdot\left( \rho_0 \mathbf{v} _1 \right) =0 \; ,
\end{equation}
and the adabatic energy equation
\begin{equation}\label{eq:2.6}
\frac{\partial p_1} {\partial t} = -\gamma p_0 \left( \nabla \cdot \mathbf{v} _1  \right) \; .
\end{equation}
We will not discuss equation (\ref{eq:2.5}) further as it can be solved fully once we know $\mathbf{v} _1$.
In this paper, we assume the background gas density is uniform and label it as $\rho_0$. A spatial variation in $\rho _0$ can cause phase mixing, e.g Heyvaerts \& Priest (\citeyear{HP1983}) and Hood {\it{et al.}} (\citeyear{Hood2002}). The phase mixing of Alfv\'en waves near a 2D magnetic null point has been looked at specifically in McLaughlin (\citeyear{M2013}).


\subsection{Coordinate system and non-dimensionalization}\label{sec:1.2b}

We now consider a coordinate system for $\mathbf{v}_1$ such that we split the velocity into  parallel, ${\rm{v}}_\parallel$, and perpendicular, ${\rm{v}}_\perp$, components. This will make our MHD mode interpretation and detection easier later, e.g. since a low-$\beta$ slow wave is wave-guided and therefore will appear primarily in ${\rm{v}}_\perp$. Thus, we let
\begin{eqnarray*}
\mathbf{v}_1 = {V}_\parallel \left( \frac {\mathbf{B}_0 } {\sqrt {\mathbf{B}_0 \cdot \mathbf{B}_0} } \right) + {V}_\perp \left( \frac { - \nabla A_0 } {\sqrt {\mathbf{B}_0 \cdot \mathbf{B}_0} } \right) + v_y \: {\hat{\bf{y}}}
\end{eqnarray*}
where $A_0$ is the vector potential and the terms in brackets are unit vectors. To aid the numerical calculation, our primary variables are considered to be ${\rm{v}}_\perp = \sqrt {\mathbf{B}_0 \cdot \mathbf{B}_0} {V}_\perp$ and ${\rm{v}}_\parallel = \sqrt {\mathbf{B}_0 \cdot \mathbf{B}_0} V_\parallel$.


We now consider a change of scale to non-dimensionalize: let ${\rm{\bf{v}}}_1 = \bar{\rm{v}} {\mathbf{v}}_1^*$, ${\rm{v}}_\perp =  \bar{\rm{v}} B {\rm{v}}_\perp^*$,${\rm{v}}_\parallel =  \bar{\rm{v}} B {\rm{v}}_\parallel^*$,  ${\mathbf{B}}_0 = B {\mathbf{B}}_0^*$, ${\mathbf{b}} = B {\mathbf{b}}^*$, $x = L x^*$, $z=Lz^*$, $p_1 = p_0 p_1^*$,       $\nabla = \nabla^* / L$ , $t=\bar{t}t^*$, $A_0=B L A_0^*$ and $\eta = \eta_0$, where we let * denote a dimensionless quantity and $\bar{\rm{v}}$, $B$, $L$, $p_0$, $\bar{t}$ and $\eta_0$ are constants with the dimensions of the variable they are scaling. We then set $ {B} / {\sqrt{\mu \rho _0 } } =\bar{\rm{v}}$ and $\bar{\rm{v}} =  L  / {\bar{t}}$; this sets $\bar{\rm{v}}$ as a constant equilibrium Alfv\'{e}n speed. We also set $ {\eta_0 \bar{t} } / {L^2} =R_m^{-1}$, where $R_m$ is the magnetic Reynolds number, and set $ {\beta_0} =  {2 \mu p_0} / {B^2}$, where $\beta_0$ is the plasma-$\beta$ at a distance unity from the origin; see also $\S\ref{sec:5.3}$.

This process non-dimensionalizes equations (\ref{eq:2.3}) - (\ref{eq:2.6}) and under these scalings $t^*=1$ refers to $t=\bar{t}=  {L} / {\bar{\rm{v}}}$; i.e. the time taken to travel a distance $L$ at the equilibrium Alfv\'en speed. {{For example, for typical coronal parameters of, say, $ \bar{\rm{v}} = 1000$\:km/s (for fast waves) and $L=1$\:Mm gives  ${\bar{t}}=  {L} / {\bar{\rm{v}}} = 1$\:second}}. For the rest of this paper, we drop the star indices; the fact that they are now non-dimensionalized is understood.


\subsection{Linearized equations}\label{sec:1.3}

Implementing our choice of coordinate system ($\S\ref{sec:1.2b}$), equations (\ref{eq:2.3}) - (\ref{eq:2.6}) become
\begin{eqnarray}
\rho_0 \frac{\partial }{\partial t}  {{\rm{v}_\perp}} &=& - \left( \mathbf{B}_0 \cdot  \mathbf{B}_0 \right)\left( \frac{\nabla \times  \mathbf{b}}{\mu} \right) +   \nabla A_0 \cdot \nabla p_1  \nonumber  \; \\
\rho_0 \frac{\partial }{\partial t}  {{\rm{v}_\parallel}}  &=& -  \left( \mathbf{B}_0 \cdot \nabla \right) p_1          \nonumber       \;\\
\rho_0 \frac{\partial }{\partial t}  v_y  &=&  \;\; \left(  \mathbf{B}_0 \cdot  \nabla \right) b_y \;   \nonumber \\
\frac{\partial  }{\partial t} {b}_x &=& -\frac{\partial }{\partial z}    {{\rm{v}_\perp}}    + \frac{1}{R_m} \nabla^2{{b}}_x     \nonumber     \; \\
\frac{\partial  }{\partial t} {b}_y &=& \left(  \mathbf{B}_0 \cdot  \nabla \right) v_y   + \frac{1}{R_m} \nabla^2{{b}}_y     \nonumber     \; \\
\frac{\partial  }{\partial t} {b}_z &=& \;\:\: \frac{\partial }{\partial x}    {{\rm{v}_\perp}}  + \frac{1}{R_m} \nabla^2 {{b}}_z     \nonumber     \; \\
\frac{\partial  }{\partial t} p_1&=&   -\gamma p_0 \left[ \nabla \cdot \left( \frac {\mathbf{B}_0 {{\rm{v}_\parallel}}}{\mathbf{B}_0 \cdot \mathbf{B}_0 }\right) - \nabla  \cdot  \left( \frac {{{\rm{v}_\perp}} \nabla A_0}{\mathbf{B}_0 \cdot \mathbf{B}_0 }\right) \right]  \label{finitebetaequations_2}\;\;. 
\end{eqnarray}
Note that in this geometry, the linearized MHD equations naturally decouple into two sets of equations: one  for the magnetoacoustics  waves and another for the Alfv\'en wave. In other words, the $y-$components of ${{\bf{v}}_1}$ and ${\bf{b}}$ (namely $v_y$ and $b_y$) entirely decouple from the  $x-$ and $z-$components. The behaviour of the Alfv\'en wave has already been investigated in McLaughlin (\citeyear{M2013}) and so we do not consider these $y-$components further: we can set $v_y=b_y=0$ without any loss of generality.

We substitute in the form of our equilibrium magnetic field (equation \ref{eq:1.2}) and apply our non-dimensionalization from $\S\ref{sec:1.2b}$, e.g. now ${\bf{B}}_0=(x,0,-z)$ and $A_0=-xz$, {{where ${\bf{B}}_0= \nabla \times {{ A_0 \: {\hat{\bf{y}} } }}$.}}    We also assume that the background gas density is uniform and so $\rho_0=1$ in our non-dimensionalized units. This gives our  linearized, non-dimensionalized perturbation equations with pressure and resistivity included. These are
\begin{eqnarray}
 \frac{\partial }{\partial t} {{\rm{v}_\perp}}  &=& v_A^2 \left( x,z \right) \left( \frac{\partial b_z}{\partial x} - \frac{\partial b_x}{\partial z}  \right) - \frac {\beta_0}{2} \left( z \frac{ \partial p_1}{\partial x}  + x   \frac{ \partial p_1}{\partial z} \right)    \nonumber \\
\frac{\partial }{\partial t} {{\rm{v}_\parallel}}  &=& - \frac {\beta_0}{2}  \left( x \frac{ \partial p_1}{\partial x} -z \frac{ \partial p_1}{\partial z} \right)  \nonumber \\
\frac{\partial b_x}{\partial t} &=& -\frac{\partial       }{\partial z}  {{\rm{v}_\perp}} + \frac {1}{R_m} \left( \frac{\partial^2 b_x}{\partial x^2} + \frac{\partial^2 b_x}{\partial z^2} \right)  \nonumber\\
\frac{\partial b_z}{\partial t} &=& \; \; \;  \frac{\partial }{\partial x} {{\rm{v}_\perp}}  + \frac {1}{R_m} \left( \frac{\partial^2 b_z}{\partial x^2}+ \frac{\partial^2 b_z}{\partial z^2} \right)  \nonumber\\
\frac{\partial p_1 }{\partial t} &=&  \frac {-\gamma}{x^2+z^2} \left[ \left(  x \frac{\partial  {{\rm{v}_\parallel}} }{\partial x} - z \frac{\partial  {{\rm{v}_\parallel}} }{\partial z} \right) - 2\:\frac {x^2-z^2}{x^2+z^2} \;  {{\rm{v}_\parallel}}  \right.\nonumber\\
&+&  \quad \quad \quad \quad \left. \left( z \frac{\partial  {{\rm{v}_\perp}} }{\partial x}  + x \frac {\partial  {{\rm{v}_\perp}}  }{\partial z} \right) - \frac {4 x z} {x^2+z^2} \;  {{\rm{v}_\perp}} \; \right] \label{finitebetaequations}
\end{eqnarray}
where the (non-dimensional) Alfv\'{e}n speed, $v_A \left( x,z \right)$, is equal to $\sqrt{x^2+z^2}$. These are the equations we will be solving in the subsequent sections.


\subsection{Plasma-$\beta$}\label{sec:5.3}

\begin{figure}[t]
\begin{center}
\includegraphics[width=2.15in]{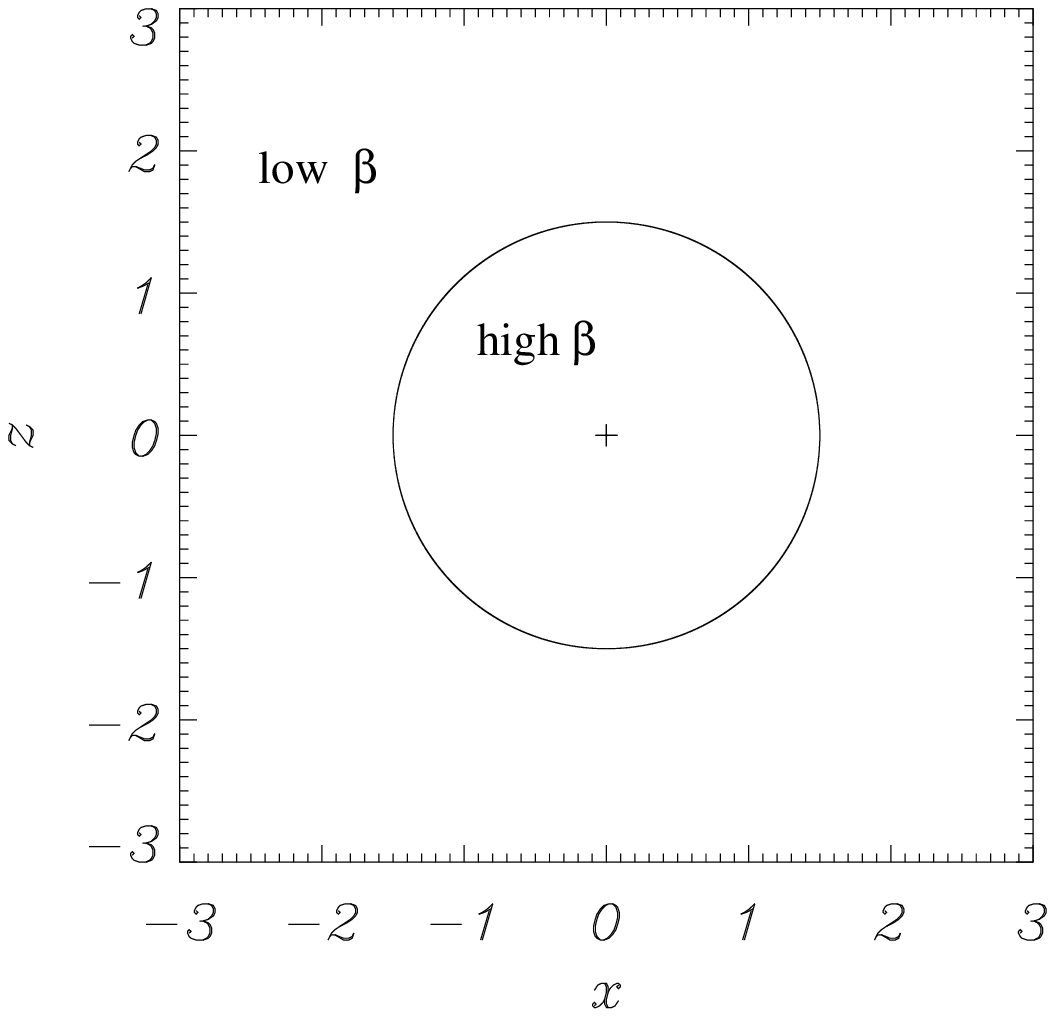}
\caption{Regions of high and low $\beta$ in our equilibrium magnetic field, where $\beta= \frac{2p_0}{x^2+z^2}$. The black circle indicates the position of the $\beta=1$ layer, where ${x^2+z^2}={\beta_0}$.}
\label{fig:areasofhighandlow}
\end{center}
\end{figure}

A parameter of key importance  in equations (\ref{finitebetaequations}) is   $ {\beta_0} =  {2 \mu p_0} / {B^2}$, where $\beta_0$ is the plasma-$\beta$ at a distance of unity from the null point. This dimensionless parameter governs the strength of the coupling between the equations for ${{\rm{v}_\perp}}$ and ${{\rm{v}_\parallel}}$. The plasma-$\beta$ parameter is defined as the ratio of the thermal plasma pressure to the magnetic pressure. In most parts of the corona, the plasma $\beta$ is much less than unity and hence the pressure gradients in the plasma can be neglected. Near magnetic null points however the magnetic field is diminishing (actually reaching zero at the null itself) and so the plasma-$\beta$ can become very large. Note that in this paper, $\beta$ denotes the true plasma-$\beta$ and $\beta_0$ denotes  the value of the  plasma-$\beta$ at a radius of unity; $r=1$.  Thus, the true plasma-$\beta$ varies through the whole region, since magnetic  field is varying everywhere throughout our model; see Figure \ref{fig:areasofhighandlow}. In our system, $\beta \propto \left({x^2+z^2} \right)^{-1}$ and thus will reach infinity at the null; here the origin. Thus, considering equilibrium quantities
\begin{eqnarray}
 \beta &=&  \frac {  {\rm{thermal \: \:plasma \:\: pressure}}}{{\rm{magnetic \:\: pressure}}}  =   \frac {p_0} {   {\left( \mathbf{B}_0 \cdot  \mathbf{B}_0 \right)}  / 2\mu} =   \frac{2 \mu p_0   / B^2}{x^2+z^2} =  \frac{\beta_0}{x^2+z^2} =\frac{\beta_0}{r^2}\;,  \label{night1}
\end{eqnarray}
where $r^2=x^2+z^2$ and so we can think of the $\beta=1$ layer as occurring at radius $r = \sqrt{\beta_0}$, i.e. this is the radius at which  the thermal plasma pressure and  magnetic pressure are equal.

There is coupling between the perpendicular and parallel velocity components specifically through $\beta_0$ and this coupling is most effective where the sound speed, $c_s$, and the Alfv\'en speed, $v_A$,  are comparable in magnitude, i.e. where  $c_s^2=v_A^2$. Bogdan {\emph{et al.}} (\citeyear{Bogdan2003}) refer to this zone the magnetic canopy or the  $\beta \approx 1$ layer. Here we define the equilibrium  sound speed as 
\begin{equation*}
c_s= \sqrt{\frac{\gamma p_0}{\rho_0}} \;\;\;{\rm{where}}\;\;  \beta_0 = {2 \mu p_0} / {B^2}  =  {2 p_0} / {{ \rho_0}   {\bar{\rm{v}}}^{\:2}}  \;    \; \Rightarrow \;\;\; c_s = \sqrt{\frac{\gamma}{2} \beta_0  } {\bar{\rm{v}}}  = \sqrt{\frac{\gamma}{2} \beta \left(x^2+z^2\right) }   {\bar{\rm{v}}} 
\end{equation*}
where we non-dimensionalize the sound speed such that $c_s = {\bar{\rm{v}}} c_s^*$ and, as before, drop the star indices.


Thus, the  $c_s^2=v_A^2$ layer occurs at:
\begin{eqnarray}
{\frac{\gamma}{2}} \beta \left(x^2+z^2\right) = x^2+z^2 \;\; \Rightarrow \;\; {\frac{\gamma}{2} \beta}=1 \quad {\rm{or\;alternatively}} \quad \frac{\gamma}{2} \beta_0 = x^2+z^2 \; \Rightarrow \;\;  r= \sqrt{\frac{\gamma     \beta_0}{2} }\label{beta_0_choice}
\end{eqnarray}
where we recall $v_A^2=x^2+z^2$ in non-dimensionalized variables. Thus, the  $c_s^2=v_A^2$ layer, or alternatively the $\beta={2}/{\gamma}$ layer, occurs at a radius $r= \sqrt{{\gamma \beta_0} / 2}$.  This is the  radius at which the Alfv\'en speed and sound speed are comparable, and it is through this that the mixing and/or coupling arises with the greatest efficiency. Of course, the difference between the $\beta=1$ layer at  $r = \sqrt{\beta_0}$  and the $c_s^2=v_A^2\:$ layer at $r= \sqrt{{\gamma \beta_0} / 2}$ is very small, and hence can be grouped together as the $\beta \approx 1$ layer. Thus, Bogdan  {\emph{et al.}} (\citeyear{Bogdan2003}) are justified in considering  the  $\beta \approx 1$ layer to be the critical layer.



\newpage

\section{$\beta=0$  magnetoacoustic wave propagation}\label{sec:2.3.3}

In this section, we begin our investigation under the $\beta=0$ assumption; equivalent to $\beta_0=0$. We also neglect the magnetic resistivity  ($\eta$)  but will discuss its role in the conclusions. Thus, we take $\eta=0$ which is equivalent to letting $R_m\rightarrow \infty$. This is referred to as an ideal plasma. This simplifies the governing equations  (\ref{finitebetaequations}) to the following
\begin{eqnarray}
\frac{\partial }{\partial t} {{\rm{v}_\perp}}  &=& v_A^2 \left( x,z \right) \left( \frac{\partial b_z}{\partial x} - \frac{\partial b_x}{\partial z}  \right)    \nonumber \\
\frac{\partial b_x}{\partial t} &=& -\frac{\partial       }{\partial z}  {{\rm{v}_\perp}} + \frac {1}{R_m} \left( \frac{\partial^2 b_x}{\partial x^2} + \frac{\partial^2 b_x}{\partial z^2} \right)  \nonumber\\
\frac{\partial b_z}{\partial t} &=& \; \;   \frac{\partial }{\partial x} {{\rm{v}_\perp}}  + \frac {1}{R_m} \left( \frac{\partial^2 b_z}{\partial x^2}+ \frac{\partial^2 b_z}{\partial z^2} \right)  \label{beta0equations}
\end{eqnarray}
where, as before, the Alfv\'{e}n speed $v_A \left( x,z \right)$$=$$ \sqrt{x^2+z^2}$ and  ${ \mathbf{b} } = \left( b_x,0,b_z \right)$.

Note that here  ${{\rm{v}_\parallel}}={\rm{constant}}$ and so, if initially absent, the slow magnetoacoustic wave is always absent under the $\beta=0$ assumption; as expected. 

We note that these equations can now be combined to form a single wave equation with a spatially-varying speed 
\begin{eqnarray}
  \frac{\partial ^2 }{\partial t^2}  {\rm{v}_\perp}=    v_A^2  \left( \frac{\partial^2 }{\partial x^2} + \frac{\partial ^2 }{\partial z^2}  \right)  {\rm{v}_\perp} =    (x^2+z^2) \left( \frac{\partial^2 }{\partial x^2} + \frac{\partial ^2 }{\partial z^2}  \right)  {\rm{v}_\perp}   =(x^2+z^2) \nabla^2    {\rm{v}_\perp}           \;               \label{fastbetachapter3}.  
\end{eqnarray}

From equation (\ref{fastbetachapter3}) is apparent that the Alfv\'en speed $v_A^2=x^2+z^2$ plays a vital role in the wave evolution. Thus, the natural choice here is to switch to polar coordinates. Other authors have looked at the behvaiour around a  null point using a Cartesian system, e.g. McLaughlin \& Hood \citeyear{MH2004}. However, changing to polar coordinates allows these equations to be examined using  analytical and semi-analytical approaches, and so may add to our understanding of such a system.

In polar coordinates, the magnetic field described by equation (\ref{eq:1.2}) and seen in Figure \ref{finitebetaXpoint} is
\begin{eqnarray}
{\bf{B}}_0 = {-r \cos{2\theta} \:{\bf{\hat{r}}}} + {r \sin{2\theta}\:} {  { \bf{{\hat\theta}} }  }   \label{polarNULL}
\end{eqnarray}
Thus
\begin{eqnarray}
{\bf{B}}_0 \cdot {\bf{B}}_0=r^2 \;\; , \quad \nabla \times {\bf{b}} = \frac{1}{r}\left[ \frac{\partial}{\partial r} \left( r b_{\theta} \right) - \frac{\partial}{\partial \theta} b_r \right] {\hat{\bf{y}}}\;\; , \quad {\bf{A}} =-{\frac{1}{2}} r^2 \sin{2\theta}  \: {\hat{\bf{y}}}\label{polarNULL2} \;.
\end{eqnarray}
Here, the linearised equations for the $\beta=0$ fast magnetoacoustic wave, i.e. the non-dimensionalized equivalents of equations (\ref{beta0equations}), are
\begin{eqnarray*}
  \frac{\partial {\rm{v}}_\perp}{\partial t} = r^2 \left[ \frac{1}{r} \frac{\partial}{\partial \theta} b_r -  \frac{1}{r} \frac{\partial}{\partial r} \left( r b_{\theta} \right) \right] \;, \qquad \frac{\partial b_r}{\partial t} = {\frac{1}{r}}\frac{\partial {\rm{v}}_\perp}{\partial \theta} \;, \qquad \frac{\partial b_\theta}{\partial t} =   -\frac{\partial {\rm{v}}_\perp}{\partial r} \;\label{fastalphapolar}.
\end{eqnarray*}
As in equation (\ref{fastbetachapter3}), these can be combined to form a single wave equation:
\begin{eqnarray}
\frac{\partial^2 \rm{v}_\perp}{\partial t^2}      &=& r^2 \left[ \frac{1}{r^2} \frac{\partial^2 \rm{v}_\perp  }{\partial \theta^2}  +  \frac{1}{r} \frac{\partial}{\partial r} \left(  r \frac{\partial {\rm{v}_\perp}}{\partial r} \right) \right]= r^2 \nabla^2 {\rm{v}}_\perp   \label{fastbetapolar}
\end{eqnarray}
where we have used the polar coordinates form of $\nabla^2=\frac{1}{r}\frac {\partial}{\partial r} \left( r \frac {\partial}{\partial r}   \right) + \frac{1}{r^2}\frac {\partial^2 }{\partial \theta^2 }$.  Note that we can change between equations (\ref{fastbetachapter3}) and (\ref{fastbetapolar}) using the substitution $x=r\cos{\theta}$, $z=r\sin{\theta}$ and $r^2=x^2+z^2$.


\subsection{Analytical solution: Klein-Gordon and Bessel functions}\label{sec:2.3.3.b}

Equation (\ref{fastbetapolar}) is a 2D wave equation with an equilibrium  Alfv\'{e}n speed that is spatially varying. Since it is a wave equation, we would expect to proceed by the usual Fourier component substitution. However, we are unable to do this here because we do not have constant coefficents in equation (\ref{fastbetapolar}). Instead we shall perform some mathematical manipulation. The right-hand-side of  equation          (\ref{fastbetapolar})      is $  r \frac {\partial }{\partial r} \left( r \frac {\partial {\rm{v}_\perp} }{\partial r}   \right) + \frac {\partial ^2 {\rm{v}_\perp} }{\partial \theta ^2 } $ and we can proceed by considering a change of variable. Let $u= \ln {r} -\ln{r_0} $ where ${r}/{r_0}$ is a dimensionless quantity. Thus equation (\ref{fastbetapolar}) becomes
\begin{eqnarray}
\frac {\partial^2 {\rm{v}}_\perp } {\partial t^2} = r \frac {\partial }{\partial r} \left( r \frac {\partial {\rm{v}_\perp} }{\partial r}   \right) + \frac {\partial ^2 {\rm{v}_\perp} }{\partial \theta ^2 } = \frac {\partial }{\partial u} \left( \frac {\partial{\rm{v}}_\perp }{\partial u}   \right) + \frac {\partial^2{\rm{v}}_\perp }{\partial \theta^2 } = \frac {\partial^2 {\rm{v}}_\perp}{\partial u^2}  + \frac {\partial^2 {\rm{v}}_\perp}{\partial \theta^2 } \; \;,\label{XYZ}
\end{eqnarray}
where $\frac{d u} {d r} = \frac {1} {r}$ and so $\frac {\partial} {\partial u } = \frac {d r} {d u} \cdot \frac {\partial } {\partial  r } = r \frac {\partial } {\partial  r } $. Note this substitution works equally well for $u=+\ln {\frac{r}{r_0}} $ or $u=-\ln {\frac{r}{r_0}} $ since the signs just cancel out. Here $r_0$ is an imposed constant and has the effect of setting $u=0$ at $r=r_0$ and so $r_0$ can be thought of as a boundary. This is discussed further in $\S\ref{concl_1}$.

Using this substitution, we now have constant coefficients. Typically we would now try a harmonic solution such that $ {\rm{v}}_\perp= e^{i \left( \omega t + n u + m \theta \right) }$ and this would give a dispersion relation via normal mode analysis. However, we have to be careful as  $n$ may be complex due to our substitution. In fact, the only separable part we can really justify is that the $\theta$-dependence satisfies $ \sim e^{i m \theta} $ so that we have periodicity, where $m$ is an integer and represents the   azimuthal wavenumber.

We now assume we can separate variables such that ${\rm{v}}_\perp \left( u,t,\theta \right) = \sigma \left( u,t \right) \cdot \Theta \left( \theta \right) $. So
\begin{eqnarray*}
 \Theta  \frac {\partial^2 \sigma } {\partial t^2} - \Theta \frac {\partial^2 \sigma}{\partial u^2 }            &=& \sigma  \frac {\partial^2  \Theta  }{\partial \theta^2} \\
\Longrightarrow  \quad\frac {\sigma_{tt} } {\sigma} - \frac{{\sigma_{uu} }}{\sigma} &=& \frac { \Theta_{\theta\theta} } {\Theta} = \textrm{constant} = -m^2\\
\Longrightarrow \qquad \quad \;\: \Theta_{\theta\theta} &=& -m^2 \Theta \quad \Longrightarrow \quad \Theta \left( \theta \right) = A\cos {m\theta} +  B\sin {m\theta} \; ,
\end{eqnarray*}
where $A$ and $B$ are constants. Thus $\frac {\partial^2  {\rm{v}}_\perp      } {\partial \theta^2}= \Theta_{\theta\theta}\sigma = -m^2 \Theta \sigma = -m^2 {\rm{v}}_\perp $ and so  equation (\ref{XYZ}) simplifies to
\begin{eqnarray}
\frac {\partial^2  {\rm{v}}_\perp } {\partial t^2} &=& \frac {\partial^2  {\rm{v}}_\perp}{\partial u^2}  - m^2  {\rm{v}}_\perp \qquad\Longrightarrow \qquad   \frac {\partial^2 \sigma } {\partial t^2} = \frac {\partial^2 \sigma}{\partial u^2}  - m^2 \sigma \; .
\label{kleingordon}
\end{eqnarray}
We identify this equation as a {\emph{Klein-Gordon}} equation.


\subsubsection{Klein-Gordon with $m=0$}\label{sec:2.3.3.b.1}

The Klein-Gordon equation (\ref{kleingordon}) is a modified wave equation and it can be solved  analytically. Firstly, we look at the simplest solution where $m=0$. Setting $m=0$ reduces the Klein-Gordon equation to the familiar wave equation $\frac {\partial^2 \sigma } {\partial t^2} = \frac {\partial^2  \sigma } {\partial u^2 } $. This has a D'Alembert solution and so
\begin{equation*}
\sigma = \mathcal{F} \left( u - t  \right) + \mathcal{G} \left( u + t  \right) \; , 
\end{equation*}
where $\mathcal{F}$ and $\mathcal{G}$ are arbitrary functions determined by the initial and boundary conditions. Note the arguments are dimensionless. Using our substitutions  $u= \ln { \frac {r} {r_0} } $ where $r^2=x^2+z^2$ and  $r_0^2=x_0^2+z_0^2$, and recalling that $\Theta = A\cos {m \theta} + B \sin {m \theta}$ and so  $\Theta$ is a constant for  $m=0$  that we can absorb into the arbitrary functions, gives
\begin{eqnarray}
{{\rm{v}_\perp}} (u,t) &=&\mathcal{F} \left[  \frac {1} {2} \log { \left( \frac {r^2} {r_0^2}  \right)}  - t \right] \quad\quad+\mathcal{G} \left[ \frac {1} {2} \log { \left( \frac {r^2} {r_0^2} \right) }  + t  \right] \nonumber \\
&=& \mathcal{F} \left[  \frac {1} {2} \log { \left( \frac {x^2+z^2} {x_0^2+z_0^2}  \right)}  - t \right] + \mathcal{G} \left[\frac {1} {2} \log { \left( \frac {x^2+z^2} {x_0^2+z_0^2} \right) }  + t  \right] \nonumber \\ 
 &=&\mathcal{F} \left[ \pm   \log { \left( \frac {r} {r_0}  \right)}  - t \right] \quad\quad+\mathcal{G} \left[  \pm  \log { \left( \frac {r} {r_0} \right) }  + t  \right] \;.\label{loglog}
\end{eqnarray}


\subsubsection{Klein-Gordon $m \neq 0$}\label{sec:2.3.3.b.1.2}

We can also solve the Klein-Gordon equation for $m \neq 0$. Starting with the case $m=1$,  equation (\ref{kleingordon}) becomes $\frac {\partial^2 \sigma } {\partial t^2} = \frac {\partial^2 \sigma}{\partial u^2}  - \sigma$. Letting $s = \sqrt { t^2 - u^2 }$ gives  $ \frac {\partial} {\partial t} = \frac {t} {s} \frac {d} {d s}$ and $ \frac {\partial} {\partial u} = - \frac {u} {s} \frac {d} {d s}$ and so our equation becomes
\begin{eqnarray*}
\frac {t^2} {s^2} \frac {d^2 \sigma} {d s^2 } + \frac {1} {s} \frac {d \sigma} {d s} - \frac {t^2} {s^3} \frac {d \sigma} {d s } &=& \frac {u^2} {s^2} \frac {d^2 \sigma} {d s^2 } - \frac {1} {s} \frac {d \sigma} {d s} - \frac {u^2} {s^3} \frac {d \sigma} {d s} - \sigma \; \\
\Longrightarrow \qquad \qquad \frac {d^2 \sigma} {d s^2 } + \frac {1} {s} \frac {d \sigma} {d s} + \sigma &=& 0 \; .
\end{eqnarray*}
This is a {\emph{Bessel Equation}} of the form $\nu=0$. Thus, it has solution
\begin{eqnarray*}
\sigma = c_1 J_0 (s) + c_2 Y_0 (s)  
\end{eqnarray*}
where
\begin{eqnarray*}
J_0 (s) = \sum_{n=0}^{\infty}\frac { (-1)^n \cdot s^{2n} } { 2^{2n} \cdot (n!)^2 }  \quad {\rm{and}} \quad  Y_0(s) = \frac {2} {\pi} \left[ J_0(s) \cdot \left( \log {\frac {s} {2}} + \Gamma \right)  + \sum_{n=1}^{\infty}\frac { (-1)^{n+1} \cdot h_n \cdot s^{2n} } { 2^{2n} \cdot (n!)^2 } \right]  \; , 
\end{eqnarray*}
where $\Gamma$ is the Euler-Mascheroni constant, $\Gamma = \lim\limits_{x \to \infty} \left( h_m - \log {x} \right) $ and $h_m$ is the harmonic number such that $h_m = \sum\limits_{k=1}^{m}    k^{-1}  $. 

Our parameter $s$ is valid for $ t \geq u$ and so $s=0$ is allowed, thus we discount our $Y_0$ solution, due to its logarithmic term. Hence $\sigma = c_1J_0 (s)$. Now ${\rm{v}_\perp} = \sigma  \Theta $ and so our solution is
\begin{eqnarray*}
{\rm{v}_\perp} = J_0 (s) \cdot \left( A\cos {\theta} + B \sin {\theta} \right)  \; , 
\end{eqnarray*}
where we have absorbed the constant $c_1$ into $A$ and $B$. Substituting $s$ back to the original variables gives
\begin{eqnarray*}
{{\rm{v}_\perp}} = J_0 \left( {\sqrt{  t^2 - \left( \ln {\frac {r} {r_0} } \right)^2 }} \right)  \cdot \left( A\cos {\theta} + B \sin {\theta} \right) \; .
\end{eqnarray*}
This can easily be extended to the case $m \neq 0$ or 1 by rescaling $t$ and $u$. Thus, our  general $ m \neq 0$ form for ${\rm{v}_\perp}$ is
\begin{eqnarray}
{{\rm{v}_\perp}} = J_0 \left( {m \sqrt{ t^2  - \left( \ln {\frac {r} {r_0} } \right)^2  }} \right)  \cdot \left( A\cos {m \theta} + B \sin {m \theta} \right) \; \label{solutuion}.
\end{eqnarray}

Thus, in this section we have solved the Klein-Gordon equation analytically and in doing so found an analytical solution to our $\beta=0$ fast magnetoacoustic wave equation. Note that  a great deal of work has been caried out on the Klein-Gordon equation; e.g. Lamb (1909; 1932) worked with this equation whilst looking at the behaviour of sound waves. However, through making certain substitutions we have actually solved the equation for a particular solution only, i.e. an initial condition of the form $\delta \left( r-r_0\right)$. To solve the Klein-Gordon in general, we would need to use numerical techniques. Thus, in $\S\ref{sec:2.3.3.a}$ we will consider a numerical solution of our system.


\subsection{Numerical simulation}\label{sec:2.3.3.a}

Equation (\ref{fastalphapolar}) can be solved  with a number of numerical schemes with the variables defined in polar coordinates. However, polar coordinate systems have a fundamental problem when it comes to crossing the origin; firstly, the radial coordinate decreases  to zero, then increases from zero. This movement through zero also  causes an instantaneous shift of $\pi$ in the angular coordinate. This can cause a problem in many numerical codes. Secondly, dividing by $r=0$ is always a problem.

Hence, instead of utilising a 2D polar coordinates numerical code to solve equations (\ref{fastalphapolar}) where the wave is driven on the (now circular) boundary, we utilised the Cartesian, two-step Lax-Wendroff numerical scheme detailed in McLaughlin (\citeyear{M2013}) with an {\emph{initial pulse condition}} as oppposed to a driven boundary. The numerical scheme was run in a  box with $-6 \le x \le 6$ and $-6 \le z \le 6$ and an initial pulse was set up around $r=3$ such that
\begin{eqnarray}
{\rm{v}_\perp}(r, \theta,t=0) = \sqrt{3} \sin \left[ \pi \left( r-2.5 \right) \right]   \left\{\begin{array}{lc}
{\mathrm{for} \; \;2.5 \leq r \leq 3.5 }  \label{pulse_pulse} \\
{ {\mathrm{for}} \; \;\;\; 0 \leq \theta \leq 2\pi }         
\end{array} \right. \; .
\end{eqnarray}
Of course, this pulse was written into the code in terms of $x=r \cos \theta$ and $z=r \cos \theta$ so what was actually solved was
\begin{eqnarray*}
{\rm{v}_\perp}(x, z,t=0) = \sqrt{3} \sin \left[ \pi \left( {\sqrt{x^2+z^2}}-2.5 \right) \right]  \quad {\mathrm{for} \; \;2.5 \leq {\sqrt{x^2+z^2}} \leq 3.5 } \; ,\\
\left.\frac {\partial } {\partial x } {\rm{v}_\perp} \right|  _{x=-6} = 0 \; , \quad\left.\frac {\partial  } {\partial x }  {\rm{v}_\perp} \right|  _{x=6} = 0 \; , \quad\left.\frac {\partial  } {\partial z }  {\rm{v}_\perp} \right|  _{z=-6}  = 0 \; , \quad \left.\frac {\partial  } {\partial z }  {\rm{v}_\perp} \right|  _{z=6}  = 0 \; .
\end{eqnarray*}
This gave a suitable initial pulse.

\begin{figure}[t]
\begin{center}
\includegraphics[width=5.0in]{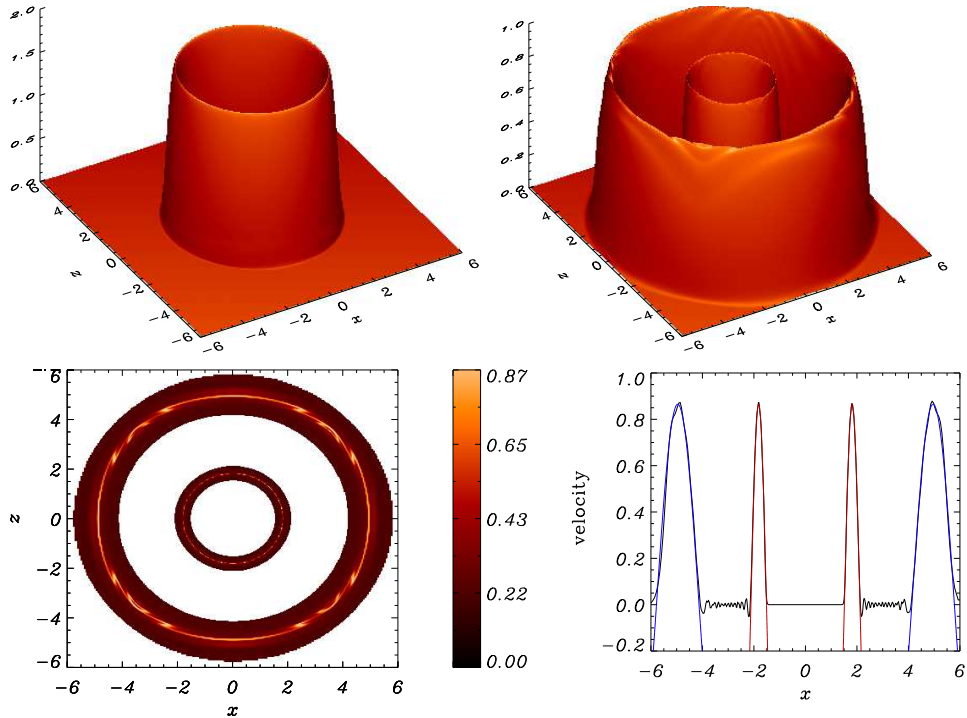}
\caption{The top left hand shaded surface shows the intial pulse at $t=0$. The top right subfigure shows the pulse split after $t=0.5$. Lower left hand side shows pulse again after $t=0.5$ but from above. Bottom right shows a cut along ${\rm{v}_\perp}\left(x,z=0\right)$ with the black line showing the numerical solution and the coloured lines showing the analytical agreement.}
\label{fig:2.3.3.1}
\end{center}
\end{figure}

When the numerical experiment began, the initial condition pulse split into two waves; each propagating  in different directions. The waves split apart naturally  and we then concentrate our attention on the incoming circular wave. The outgoing wave is not of primary concern to us and the boundary conditions let the wave pass out of the box. This can be seen in Figure \ref{fig:2.3.3.1}. The top left hand shaded surface shows the intial pulse at $t=0$. The top right subfigure shows the pulse split into two after $t=0.5$. The lower left hand side shows the pulse again after $t=0.5$ but from above. We see that the two waves have disassociated in the sense that we are free to just concentrate on the incoming solution. The bottom right subfigure will be discussed below.

We can also understand the splitting of the initial condition into two wave pulses in terms of D'Alembert's solution, as discussed in $\S\ref{sec:2.3.3.b.1}$. Here our initial condition has the form
\begin{eqnarray*}
 \sqrt{3} \sin \left[ \pi \left( r-2.5 \right) \right] &=& {\frac{1}{2}}\mathcal{F}\left(t+\log{r}\right) + {\frac{1}{2}}\mathcal{F}\left(t-\log{r}\right)\\
&=& {       \frac{\sqrt{3}}{2}} \sin \left[ \pi \left( e^{t+\log{r}} -2.5 \right) \right] + \frac{\sqrt{3}}{2} \sin \left[ \pi \left( e^{-{t+\log{r}}} -2.5 \right) \right] 
\end{eqnarray*}
These analytical descriptions match the evolution of the two waves satisfactorially and the agreement can be seen in  the bottom right subfigure of Figure \ref{fig:2.3.3.1}. {{Note how the numerical solution has some small dispersion as the two waves split; this is due to our choice of pulse (\ref{pulse_pulse}) having discontinuities in the first-derivative at its edges.}}

\begin{figure}[t]
\begin{center}
\includegraphics[width=6.0in]{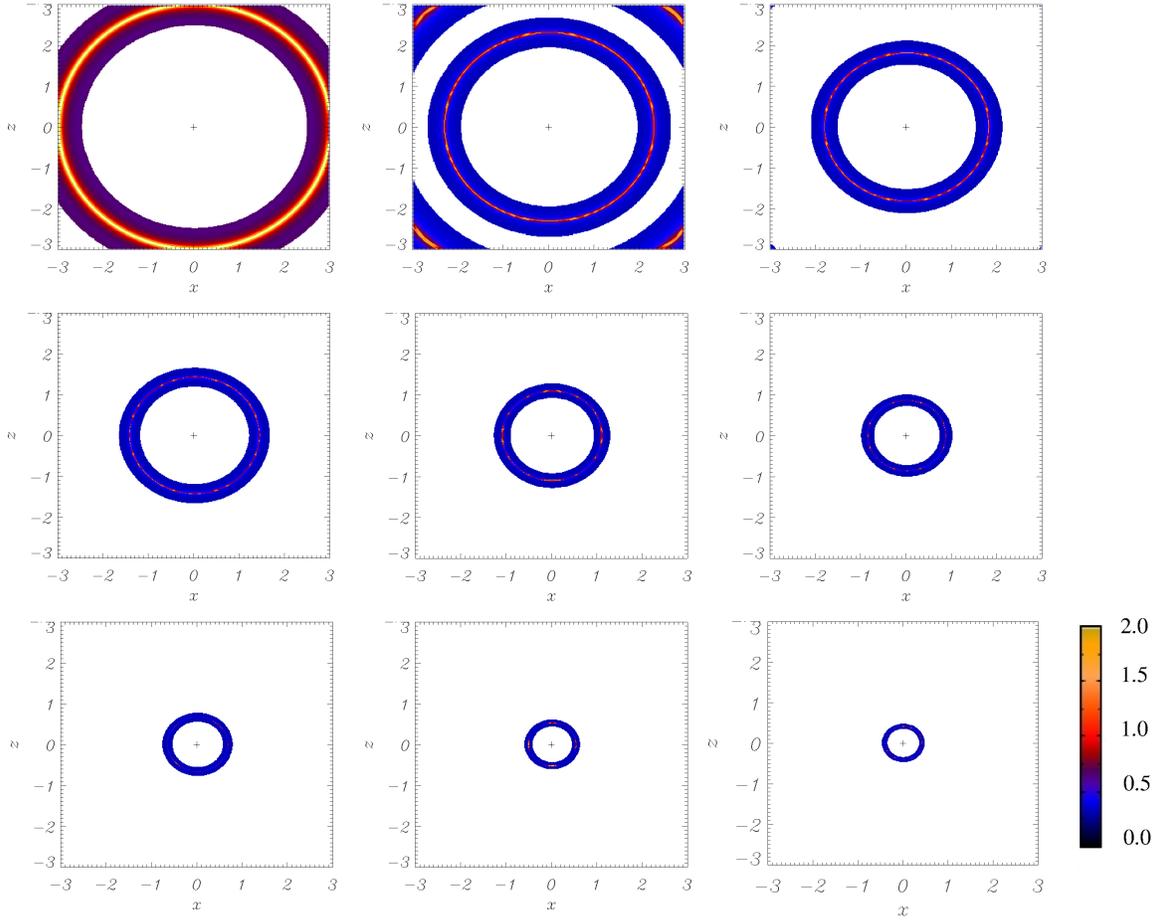}
\caption{Contours of numerical simulation of $\rm{v}_\perp$ for a fast wave pulse initially located  about a radius $\sqrt{x^2+z^2}=3$, and its resultant propagation at times $(a)$ $t$=0, $(b)$ $t$=0.25, $(c)$ $t$=0.5, $(d)$ $t=$0.75, $(e)$ $t$=1.0 and $(f)$ $t$=1.25, $(g)$ $t$=1.5, $(h)$ $t$=1.75 and $(i)$ $t$=2.0, labelling from top left to bottom right. The black cross indicates the location of the null point (at the origin). }
\label{fig:2.3.3.2}
\end{center}
\end{figure}

The simulation was run with a resolution of $1000 \times 1000$ points and successful convergence tests were performed. However, since we expect the important/interesting behaviour to occur close to the origin/null, a {{stretched grid}} was implemented to focus the majority of the grid points close to the origin. The stretching algorithm smoothly stretched the grid such that 50\% of the grid points lay within a radius of $1.5$. This gave a better resolution in the area of prime interest. The behaviour of the fast wave with a circular geometry can be seen in Figure \ref{fig:2.3.3.2}.  Note how the initial pulse can be seen in the top left subfigure and that it has magnitude $\sqrt{3}$, then at a later time the wave has split and has magnitude $ {\sqrt{3}} / {2}$.

\newpage

\subsection{Semi-analytical approach: WKB approximation}\label{sec:2.3.3.c}

We can also solve equation (\ref{fastbetapolar}) using the WKB approximation. The WKB approximation is an asymptotic approximation technique which can be used when a system contains a large parameter.  It is named after \emph{Wentzel}, \emph{Kramers} and \emph{Brillouin}, who pioneered its use in quantum mechanics around 1927. Details of the theory  can be found in Murray (\citeyear{Murray}), Sneddon (\citeyear{Sneddon}), Bender \& Orszag (\citeyear{Bender})  and Evans, Blackledge \& Yardley (\citeyear{Evans}).


 Substituting ${{\rm{v}_\perp}} = e^{i \phi (r,\theta) } \cdot e^{-i \omega t}$ into equation (\ref{fastbetapolar}) gives   
\begin{eqnarray*}
-\omega ^ 2 = \left[-r^2 \left( \frac {\partial \phi} {\partial r} \right)^2 - \left( \frac {\partial \phi} {\partial \theta} \right)^2\right] +i \left[ r^2 \left( \frac {\partial^2 \phi} {\partial r^2} \right) + r \left( \frac {\partial \phi} {\partial r} \right) + \left( \frac {\partial^2 \phi} {\partial \theta^2} \right)\right] \;. 
\end{eqnarray*}
Now we make the WKB approximation such that $\phi \sim \omega \gg 1 $, which yields
\begin{eqnarray*}
\omega ^2 = r^2p^2 + q^2 
\end{eqnarray*}
where  $p=\frac {\partial \phi} {\partial r}$ and $q=\frac {\partial \phi} {\partial \theta}$. 
This leads to the construction of a first-order, non-linear partial differential equation of the form $\mathcal{G} \left( r,\theta,\phi,p,q \right)=0$ such that
\begin{eqnarray*}
\mathcal{G} \left( r,\theta,\phi,p,q \right) = {\frac{1}{2}} \left(r^2p^2 + q^2   - \omega ^2 \right)=0 \;.
\end{eqnarray*}
Note all the imaginary terms have disappeared. We choose to introduce $1/2$ into the construction of $\mathcal{G}$ to make the equations simplify later. 

We can now apply the {\emph{Method of Characteristics}} to solve this first-order, non-linear partial differential equation. This gives 
\begin{eqnarray*}
\frac {\partial \mathcal{G} }{\partial \phi} = 0\; , \quad \frac {\partial \mathcal{G} }{\partial p} = r^2p \; , \quad \frac {\partial \mathcal{G} }{\partial q} = q \:, \quad \frac {\partial \mathcal{G} }{\partial r} = rp^2  \; , \quad \frac {\partial \mathcal{G} }{\partial \theta} =0 \;.
\end{eqnarray*}
Now we can apply Charpit's Relations to solve these equations. Charpit's Relations are general characteristic equations first used by Charpit in 1784 and Lagrange in 1779, where the method is attributed to Charpit who perfected it. Applying Charpit's Relations yields
\begin{eqnarray}
\frac {d \phi }{ds} =  \omega ^2  \; , \quad   \frac {dp}{ds} =  -rp^2\; , \quad \frac {dq}{ds} = 0 \;,\quad \frac {dr}{ds} = pr^2  \; , \quad \frac {d\theta}{ds} = q  \;, \label{polar_zero_characteristics}
\end{eqnarray}
where $\omega$ is the frequency of our wave and $s$ is some parameter along the characteristic. These five ordinary differential equation can be solved using, for example, a fourth-order Runge-Kutta method. The initial conditions are
\begin{eqnarray*}
\phi_0 = 0 \;, \; \;  r(s=0)=r_0\;, \; \;  0 \leq \theta_0 \leq 2\pi \;, \; \;  p_0 = -\frac{\omega}{r_0} \;, \; \; q_0= 0\;,
\end{eqnarray*}
where $r_0$ is the radius of the boundary that the disturbance starts from and $p_0$ is negative so this disturbance propagates towards the origin, as we concluded from $\S\ref{sec:2.3.3.b.1.2}$.  We can also see that $q=q_0=0$. Finally, $\frac{d}{ds}\left(pr\right) =0$ $\Rightarrow pr=p_0r_0=-\omega$ in agreement with the form of $\mathcal{G}$.

\begin{figure}[t]
\begin{center}
\includegraphics[width=6.0in]{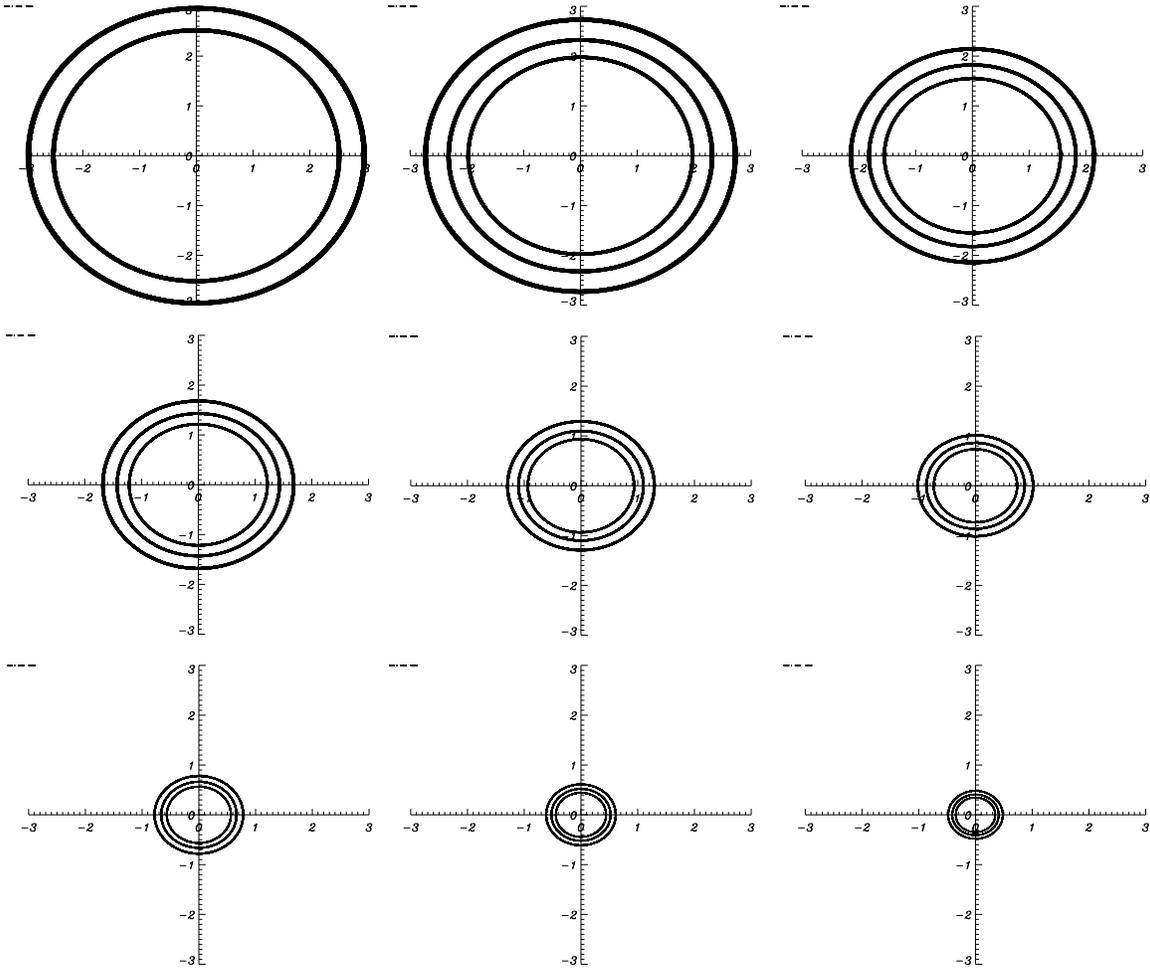}
\caption{Semi-analytical solution of ${{\rm{v}_\perp}}$ for WKB approximation of a fast wave sent in from a circular boundary at $r=2.5$, $3$ and $3.5$, and its resultant propagation at times $(a)$ $t$=0, $(b)$ $t$=0.25, $(c)$ $t$=0.5, $(d)$ $t=$0.75, $(e)$ $t$=1.0 and $(f)$ $t$=1.25, $(g)$ $t$=1.5, $(h)$ $t$=1.75 and $(i)$ $t$=2.0, labelling from top left to bottom right. The lines represent the leading, middle and trailing edges of the WKB (wave) solution.}
\label{fig:2.3.3.3}
\end{center}
\end{figure}

Thus, we can use our WKB solution to plot the evolution of the fast wave from an initial radius $r=3$ in order to compare to the numerical solution given in $\S\ref{sec:2.3.3.a}$ and Figure \ref{fig:2.3.3.2}. This can be seen in Figure \ref{fig:2.3.3.3}. The lines represent the leading, middle and trailing edges of the WKB wave solution, where the pulse starts at radii of $r=2.5$, $3$ and $3.5$.

\clearpage



\newpage

\section{$\beta \neq 0$ magnetoacoustic wave propagation}\label{polarcoorindatesbetaneq0}
 
In this section, we look at the behaviour of the fast magnetoacoustic wave in the neighbourhood of a simple 2D X-point, as we did in $\S\ref{sec:2.3.3}$. However, we now consider a $\beta \neq 0$ plasma, i.e. we lift the cold plasma restriction. This extends the model of  $\S\ref{sec:2.3.3}$  to include plasma pressure and pressure gradients and the most obvious effect of this is the introduction of {\emph{slow magnetoacoustic waves}} to the system. There will now also be the possibility of coupling between the two magnetoacoustic waves; this can be understood through the plasma-$\beta$ parameter ($\S\ref{sec:5.3}$) since there can now be an interplay between plasma pressure and magnetic pressure, and we expect this coupling and information exchange to occur primarily  near where the sound speed and Alfv\'en speed become comparable in magnitude, i.e. at the areas where the plasma-$\beta \approx 1$. Again, we will not consider the Alfv\'en wave here and recall that for Alfv\'en waves that are decoupled from fast waves, the value of the plasma-$\beta$ is unimportant since the plasma pressure plays no role in its propagation. This can also be seen mathematically in the last of equations (\ref{finitebetaequations}), i.e. neither $v_y$ nor $b_y$ appears in the equation governing $p_1$.


We approach this investigation by studying magnetoacoustic wave propagation in a  circular geometry with a similar numerical set-up to that in $\S\ref{sec:2.3.3.a}$. Again, in a circular geometry, our particular choice of magnetic field gives rise to equations (\ref{polarNULL}) and (\ref{polarNULL2}). However, we now solve the $\beta \neq 0$ linearized equations (\ref{finitebetaequations}) as opposed to the reduced $\beta=0$ set in  $\S\ref{sec:2.3.3.a}$.

There is a lot of freedom in setting $\beta_0$, where we recall  from equation (\ref{beta_0_choice}) that our choice of $\beta_0$ only affects the location of the $\beta=1$ and $c_s^2=v_A^2$ layer. This is an arbitrary choice, since our system does not have any obvious length scales. Here we choose to set $\beta_0=0.25$ and we present these  results below. We also investigated  other values of $\beta_0$ and these all give similar results; it is only the radius  of the $\beta = 1$ layer that changes in accordance with equation (\ref{beta_0_choice}). Note that for $\beta_0=0.25$,  the $\beta=1$ layer occurs at a radius $r=\sqrt{0.25}=0.5$ and correspondingly the $c_s^2=v_A^2\:$ layer occurs at a radius of $r=\sqrt{{5} / {24}}=0.456$.


As in $\S\ref{sec:2.3.3.a}$, we now solve our equations (\ref{finitebetaequations}) numerically using our 2D Cartesian Lax-Wendroff numerical code (instead of writing a polar coordinates version of the code). Thus, as before, we use the Cartesian code with an {{initial pulse condition}} and this will give us a simulation of the $\beta \neq 0$ plasma behaviour.  The numerical scheme was run in a square box with $-6 \le x \le 6$ and $-6 \le z \le 6$ and an initial pulse was set up around $r=3$ such that
\begin{eqnarray*}
{\rm{v}_\perp}(r, \theta,t=0) = \sqrt{3} \sin \left[ \pi \left( r-2.5 \right) \right]   \left\{\begin{array}{lc}
{\mathrm{for} \; \;2.5 \leq r \leq 3.5 } \\
{ {\mathrm{for}} \; \;\;\;\; 0 \leq \theta \leq 2\pi }
\end{array} \right.              \quad  {\rm{and}}  \quad    {\rm{v}_\parallel}(r, \theta,t=0) = 0             \; .
\end{eqnarray*}

Of course, this pulse was written into the code in terms of $x=r \cos \theta$ and $z=r \cos \theta$ so what was actually solved was
\begin{eqnarray}
{\rm{v}_\perp}(x, z,t=0) &=& \sqrt{3} \sin \left[ \pi \left( {\sqrt{x^2+z^2}}-2.5 \right) \right]  \quad {\mathrm{for} \; \;2.5 \leq {\sqrt{x^2+z^2}} \leq 3.5 } \; ,\nonumber\\
\left.\frac {\partial } {\partial x } {\rm{v}_\perp} \right|  _{x=-6} &=& 0 \; , \quad\left.\frac {\partial  } {\partial x }  {\rm{v}_\perp} \right|  _{x=6} = 0 \; , \quad\left.\frac {\partial  } {\partial z }  {\rm{v}_\perp} \right|  _{z=-6}  = 0 \; , \quad \left.\frac {\partial  } {\partial z }  {\rm{v}_\perp} \right|  _{z=6}  = 0 \; ,\nonumber\\
 {\rm{v}_\parallel}  (x,z,t=0) &=& 0 \; ,\nonumber\\
\left.\frac {\partial    } {\partial x }  {\rm{v}_\parallel} \right| _{x=-6} &=& 0 \; , \quad \left.\frac {\partial  } {\partial x }  {\rm{v}_\parallel} \right | _{x=6} = 0 \; , \quad\left. \frac {\partial   } {\partial z }  {\rm{v}_\parallel}  \right| _{z=6}  = 0 \; , \quad\left. \frac {\partial  {\rm{v}_\parallel}  } {\partial z }  \right| _{z=-6}  = 0\; \label{initalconditionsIC}.
\end{eqnarray}
This gave a suitable initial pulse. When the numerical experiment began, the initial condition pulse split into two waves; each propagating in different directions. The waves split naturally apart and we can then concentrate our attention on the incoming circular wave. The outgoing wave is of no concern to us and the boundary conditions let the wave pass out of the box; this concept is similar to that of $\S\ref{sec:2.3.3.a}$. The simulation was run with a resolution of $1000 \times 1000$ points, and successful convergence tests were performed. However, since we knew the important behaviour would occur close to the origin, a {{stretched grid}} was used to focus the majority of the grid points close to the null point. The stretching algorithm smoothly stretched the grid such that 50\% of the grid points lay within a radius of $1.5$. This gave better resolution in the areas of interest.

{{
Note that considering a $\beta  \neq 0 $ plasma may now also introduce the entropy mode into our system (see e.g. Goedbloed \& Poedts \citeyear{GP2004}; Murawski {\it{et al.}} \citeyear{KRIS2004}). The entropy mode  is a non-propagating MHD mode and is a solution to the ideal MHD equations with zero frequency. It can be represented as a local increase/decrease in the temperature and a decrease/increase in the mass density, but with no net pressure changes. In our system, the initial velocity pulse is  generated at $r=3$, where $\beta = \beta_0 / r^2 = 0.25/3^2 = 0.028$. Thus in our system, the entropy mode, if present, cannot propagate from its initial location and  so is outside the region of interest for our investigation.

}}



\subsection{Numerical Simulation: ${\rm{v}_\perp}$}\label{ein1}

\begin{figure}[t]
\begin{center}
\includegraphics[width=6.0in]{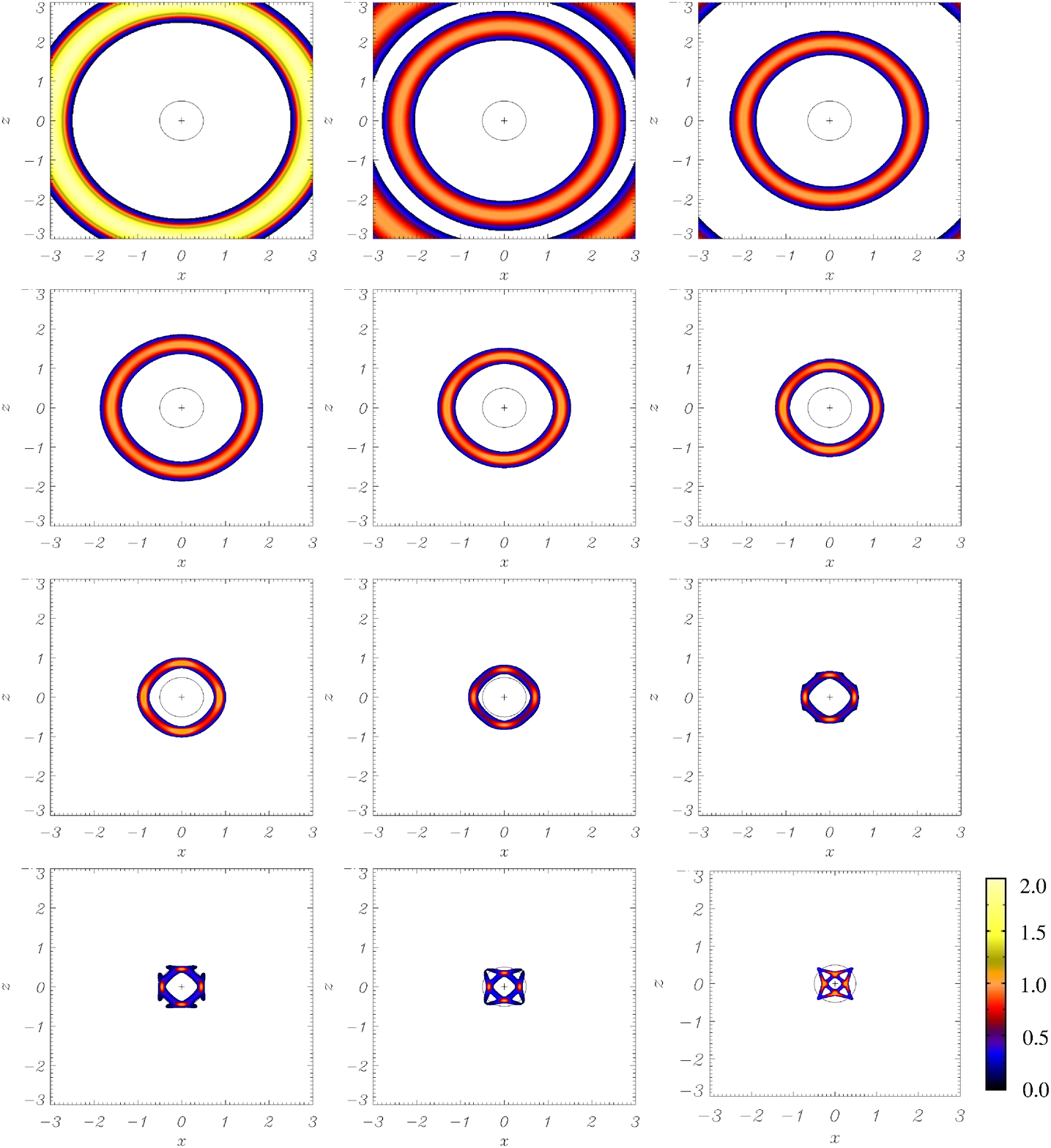}
\caption{Contours of numerical simulation of $\rm{v}_\perp$ for a fast wave pulse initially located  about a radius $\sqrt{x^2+z^2}=3$, and its resultant propagation at times $(a)$ $t$=0, $(b)$ $t$=0.2, $(c)$ $t$=0.4, $(d)$ $t=$0.6, $(e)$ $t$=0.8, $(f)$ $t$=1.0, $(g)$ $t$=1.2, $(h)$ $t$=1.4,  $(i)$ $t$=1.6, $(j)$ $t$=1.8, $(k)$ $t$=2.0 and  $(l)$ $t$=2.2, labelling from top left to bottom right. The black circle indicates the position of the  $c_s^2=v_A^2\:$ layer, which occurs at  $\sqrt{x^2+z^2}=\sqrt{\frac{\gamma \beta_0}{2}}$. The cross denotes the null point in the magnetic configuration.}
\label{fig:2.3.3.2_MAGIC}
\end{center}
\end{figure}

The evolution of the $\beta \neq 0$, linear fast magnetoacoustic wave can be seen in Figure \ref{fig:2.3.3.2_MAGIC}. We find that the fast wave splits into two waves; one approaching the origin and the other travelling away from it; as expected. The wave propagating towards the origin initially has the shape of an annulus. We find that the annulus contracts (as in $\S\ref{sec:2.3.3.a}$ and Figure \ref{fig:2.3.3.2_MAGIC}) and that, at least  initially, this contraction appeared to preserve the  original ratios (distance between the leading-and-middle wavefronts compared to middle-and-trailing wavefronts). However, as the wave continues to propagate towards the origin, it is distorted significantly  from its original shape: there is a decrease in wave  speed along the axes, i.e. the separatrices, and so the annulus starts to take on a quasi-diamond shape; with the corners located along the separatrices. This can be seen in the second and third row of subfigures of  Figure \ref{fig:2.3.3.2_MAGIC}. Eventually, the wave crosses  the  $c_s^2=v_A^2\:$ layer (indicated by a black circle in the figure, located at $r=0.456$ for $\beta_0=0.25$) where it begins a more complicated evolution: unlike that seen in the equivalent $\beta=0$ case in  Figure \ref{fig:2.3.3.2_MAGIC}.

\clearpage

\begin{figure}[t]
\begin{center}
\includegraphics[width=6.0in]{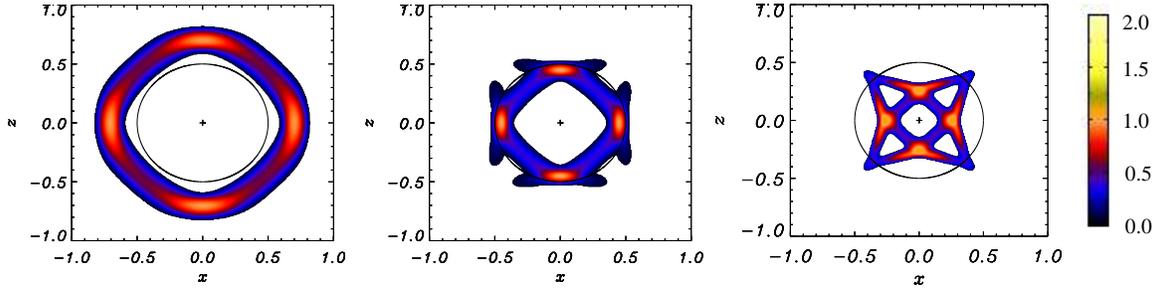}
\caption{Blow-up subfigures of  $\rm{v}_\perp$ from Figure \ref{fig:2.3.3.2_MAGIC} at times $(a)$ $t$=1.4,  $(b)$ $t$=1.8  and  $(c)$ $t$=2.2, labelling left to right.}
\label{fig:2.3.3.3_MAGIC}
\end{center}
\end{figure}

Some of the subfigures from Figure \ref{fig:2.3.3.2_MAGIC}  are shown as blown-up versions in Figure  \ref{fig:2.3.3.3_MAGIC}, specifically showing the wave evolution just before, during and just after  crossing  the   $c_s^2=v_A^2\:$ layer. 


\subsection{Numerical Simulation: ${\rm{v}_\parallel}$}\label{ein2}

\begin{figure}[t]
\begin{center}
\includegraphics[width=6.0in]{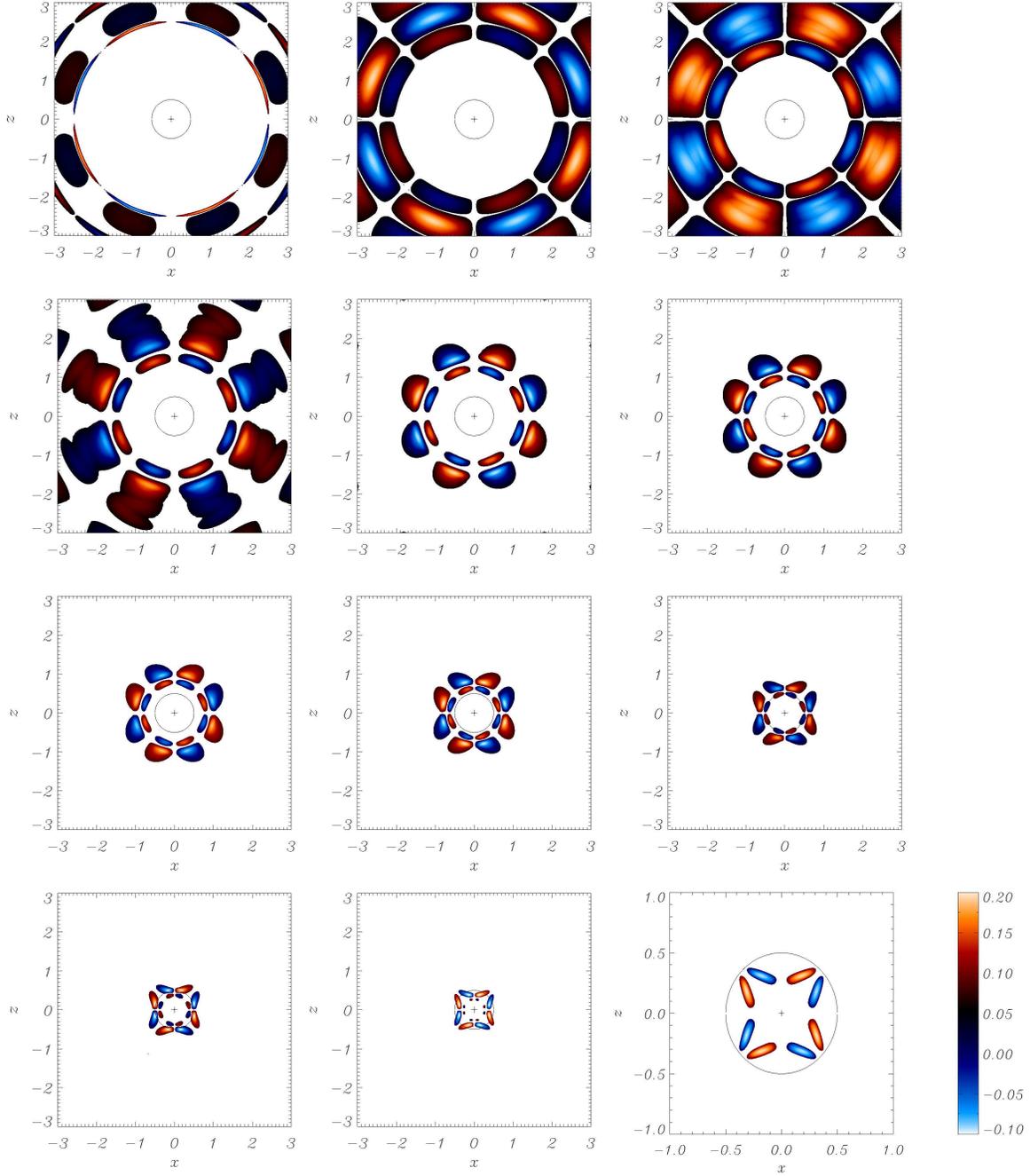}
\caption{Contours of numerical simulation of $\rm{v}_\parallel$ for a fast wave pulse initially located  about a radius $\sqrt{x^2+z^2}=3$, and its resultant propagation at times $(a)$ $t$=0.02, $(b)$ $t$=0.2, $(c)$ $t$=0.4, $(d)$ $t=$0.6, $(e)$ $t$=0.8, $(f)$ $t$=1.0, $(g)$ $t$=1.2, $(h)$ $t$=1.4,  $(i)$ $t$=1.6, $(j)$ $t$=1.8, $(k)$ $t$=2.0 and  $(l)$ $t$=2.2, labelling from top left to bottom right. The black circle indicates the position of the $\beta=1$,  ${x^2+z^2}={\beta_0}$ layer. The cross denotes the null point in the magnetic configuration. The last subfigure shows a blow-up of the central region (axes have changed).}
\label{fig:2.3.3.9}
\end{center}
\end{figure}

We can also look at the behaviour of ${\rm{v}_\parallel}$; the parallel component of our wave. This has a much more complicated behaviour than our perpendicular component and can be seen in Figure \ref{fig:2.3.3.9}. Firstly, we notice that there are both positive and negative parts to the wave, unlike the perpendicular component which was always positive. We see that the wave has an alternating structure in the $\theta$-direction. Secondly, we have set an initial condition in ${\rm{v}}_\perp$ only: the initial condition on the parallel wave was  ${\rm{v}_\parallel}  (x,z,0) = 0$ in equations (\ref{initalconditionsIC}). Hence, the ${\rm{v}_\parallel}$ wave we are observing has  been generated as a consequence of our ${\rm{v}}_\perp$ initial condition. By looking at equations (\ref{finitebetaequations}) and our initial conditions, we see that   ${\rm{v}}_\perp$  acts as a driver (forcing term) for ${\rm{v}_\parallel}$.

It is interesting to note that the waves  in Figure \ref{fig:2.3.3.9} have a smaller amplitude than those in Figure \ref{fig:2.3.3.2_MAGIC}. The ${\rm{v}_\perp}$  waves in Figure \ref{fig:2.3.3.2_MAGIC} have an amplitude of $\sim {\sqrt{3} /{2}}$ (recall the initial condition was a wave of amplitude $\sqrt{3}$ that split  in half equally) compared to the ${\rm{v}_\parallel}$ waves in Figure \ref{fig:2.3.3.9} which have an amplitude of $ \sim \beta_0 \sqrt{3} /{2}$.

\clearpage

\newpage


\section{Conclusions}\label{sec:6.10}

In this paper, we have investigated the behaviour of magnetoacoustic waves within inhomogeneous magnetic media in two specific ways:  we have investigated the behaviour of an initially cylindrically-symmetric fast magnetoacoustic wave around a 2D null point under, firstly,  the $\beta=0$  and, secondly,  the $\beta \neq 0$ assumptions.


\subsection{$\beta=0$ plasma}\label{concl_1}

In $\S\ref{sec:2.3.3}$, we  investigated the behaviour of an initially cylindrically-symmetric  fast magnetoacoustic wave around a 2D null point under the $\beta=0$ assumption. Using polar coordinates, we derived a governing wave equation with a spatially-varying characteristic speed (the Alfv\'en speed) and we solved this equation analytically by deriving a {\emph{Klein-Gordon}} equation and then solving separately for $m=0$, which led to a D'Alembert-type solution, and $m\neq 0$ which led to a Bessel-type  solution (equation \ref{solutuion}). It is interesting to note that  solution (\ref{solutuion}) is only valid for $s \geq 0$, i.e. $ t \geq \pm \ln {\frac {r} {r_0} }$, and that the same final result is gained from substituting $s = \sqrt {u^2 - t^2 }$ or  $s = \sqrt {t^2 - u^2 }$, since $J_0(s) = J_0(-s) $. We can interpret this as follows: if we consider the boundary of our system to be a shell at radius $r_0$, we can interpret the $\pm$ ambiguity on $u$ as a boundary disturbance splitting into two waves; one propagating  outwards ($r$ increasing so $r>r_0$, i.e. $u = \ln{\frac{r}{r_0}}$ solution) and one propagating  inwards ($r$ decreasing so $r<r_0$, i.e. $u = -\ln{\frac{r}{r_0}}$). Note that the inequality on $r$ here dictates the flow of information; the perturbation starts on the boundary and there is no disturbance in front of the wave, i.e. the inequality that restricts $r$ from taking certain values until time has elapsed is interpreted as regions in the system not yet  affected by the perturbation; as the information has not yet had the time to reach there since the wavefront propagates at a finite speed. Thus, if we are interested in the region inside $r=r_0$ including the origin (which is the location of our null) then we are interested primarily in the substitution $u = - \ln { \frac {r} {r_0} } $, with $r$ starting at $r_0$ and decreasing as $t$ evolves.

We also solved the $\beta=0$ governing wave equation using numerical techniques in $\S\ref{sec:2.3.3.a}$. We find that the linear, $\beta=0$ fast magnetoacoustic wave splits into two waves; one approaching the null and the other propagating away from it. The wave propagating  towards the null  has  the shape of an annulus. We find that this annulus contracts, but keeps its original ratios (distance between the leading-and-middle wavefronts compared to middle-and-trailing wavefronts). This was seen in Figure \ref{fig:2.3.3.2}. Since the Alfv\'en speed is spatially varying (i.e. $\sim r$, see equation \ref{fastbetapolar}), a \emph{refraction} effect focuses the wave into the null point. This is the same  refraction effect found in McLaughlin \& Hood (\citeyear{MH2004}). 

Finally, we investigated our system using a semi-analytical WKB approach in $\S\ref{sec:2.3.3.c}$. This can be seen in Figure \ref{fig:2.3.3.3}. As expected, the agreement between Figures \ref{fig:2.3.3.2} and \ref{fig:2.3.3.3} is excellent; the semi-analytical WKB and numerical solutions lie on top of each other. We can also see in  Figure \ref{fig:2.3.3.3}  how the ratio between the leading-and-middle and between the middle-and-trailing of the pulse is preserved. The wave focuses on the null point  and contracts around it. In addition, equations (\ref{polar_zero_characteristics}) can  be solved analytically by forming
\begin{eqnarray*}
{ \frac{dp}{ds} } \;\big{/} \;{ \frac{dr}{ds} }= \frac{dp}{dr} = -\frac{p}{r} \quad \Longrightarrow \quad  \log{r} = -\log{p} + {\rm{constant}} \quad \Longrightarrow \quad rp = -\omega \nonumber
\end{eqnarray*}
and so
\begin{eqnarray}
\frac {dp}{ds} = \omega p \; , \quad   \frac {dr}{ds} = -\omega r\; , \quad p = -\frac {\omega}{r_0} e^{\omega s} \; , \quad r = {r_0} e^{-\omega s} \label{evolution}\;,
\end{eqnarray}
where the initial conditions dictate the constants of integration. From equations (\ref{evolution}) we see $r = {r_0} e^{-\omega s}$ and so the wave, which focuses on the null and contracts around it, never actually reaches the null  in a finite time, due to the exponential decay of $r$.



\subsection{$\beta \neq 0$ plasma}

In $\S\ref{polarcoorindatesbetaneq0}$, we  investigated the behaviour of an initially cylindrically-symmetric fast magnetoacoustic wave around a 2D null point in a $\beta \neq 0$ plasma. This can be seen in Figure  \ref{fig:2.3.3.2_MAGIC}. We find that the fast wave split into two waves; one approaching the origin and the other travelling away from it; as expected. The wave propagating towards the origin initially has the shape of an annulus. We find that the annulus contracts (as in $\S\ref{sec:2.3.3.a}$ and Figure \ref{fig:2.3.3.2_MAGIC}) and that, at least  initially, this contraction appeared to preserve the  original ratios (distance between the leading-and-middle wavefronts compared to middle-and-trailing wavefronts). However, as the wave continues to propagate towards the origin, it is distorted significantly from its original shape: there is a decrease in overall wave speed along the $x=0$ and $z=0$ axes (the separatrices) and so the annulus starts to take on a quasi-diamond shape; with the corners located along the separatrices. This can be seen in the second and third row of subfigures of  Figure \ref{fig:2.3.3.2_MAGIC}. Eventually, the wave crosses  the  $c_s^2=v_A^2\:$ layer (indicated by a black circle in the figure, located at $r=0.456$ for $\beta_0=0.25$) where it begins a more complicated evolution: unlike that seen in the equivalent $\beta=0$ case in  Figure \ref{fig:2.3.3.2_MAGIC}.

\begin{figure}[t]
\begin{center}
\includegraphics[width=6.0in]{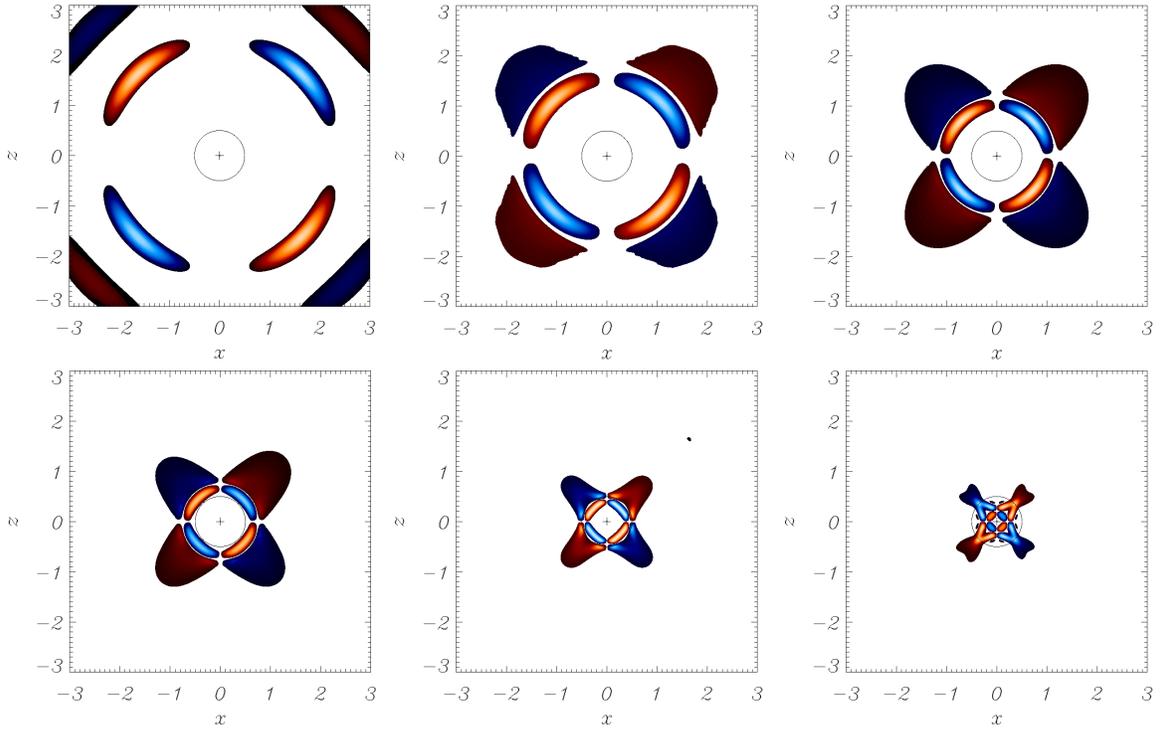}
\caption{Contours of numerical simulation of $p_1$ for a fast wave pulse initially located  about a radius $\sqrt{x^2+z^2}=3$, and its resultant propagation at times $(a)$ $t$=0.2, $(b)$ $t$=0.6, $(c)$ $t$=1.0, $(d)$ $t=$1.4, $(e)$ $t$=1.8, $(f)$ $t$=2.2, labelling from top left to bottom right. The black circle indicates the position of the  $c_s^2=v_A^2\:$ layer and the cross denotes the null point in the magnetic configuration. $p_1$ has an alternating form, where orange represents $p_1>0$ and blue represents $p_1<0$. The pressure appears to follow a $\sin{2\theta}$ pattern.}
\label{fig:2.3.3.4}
\end{center}
\end{figure}

\begin{figure}[h]
\begin{center}
\includegraphics[width=2.0in]{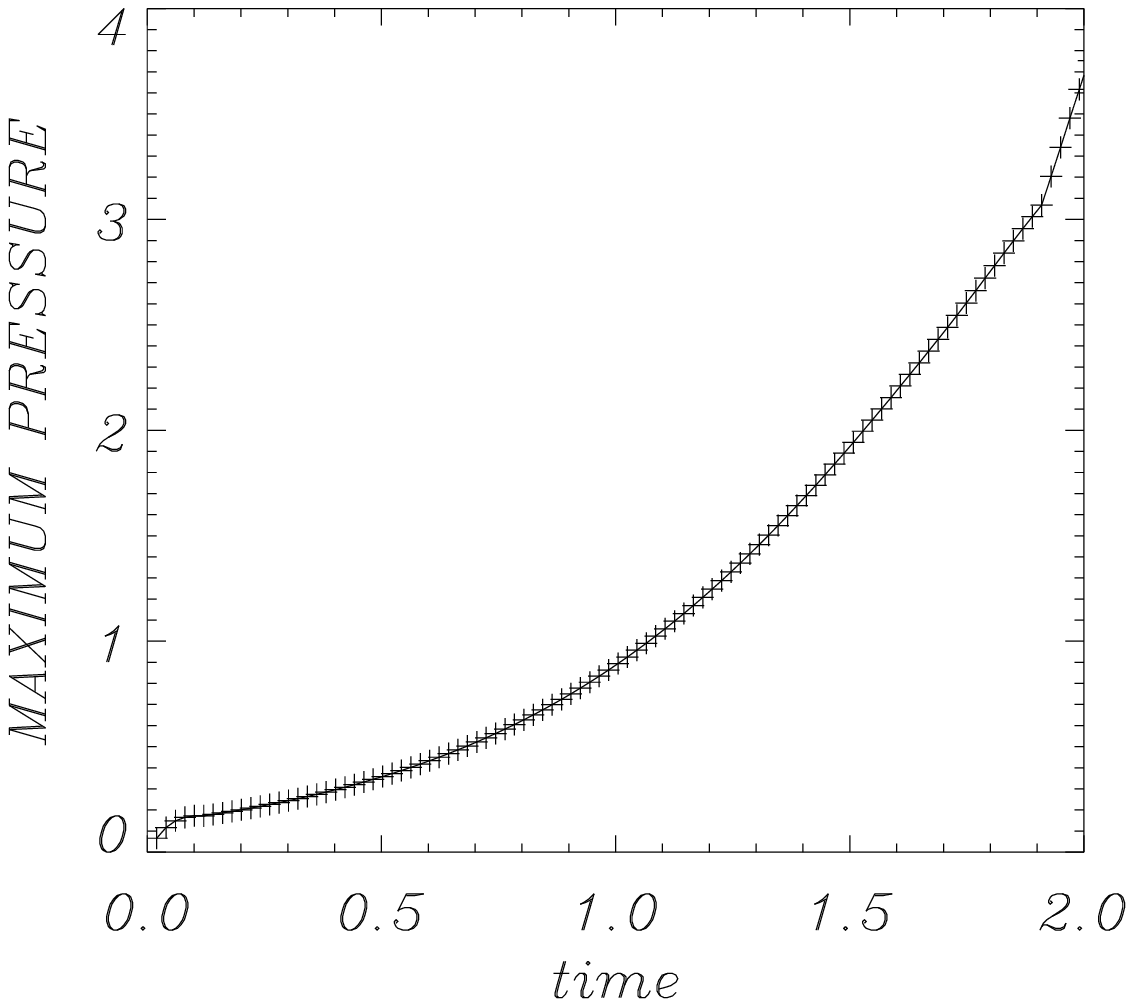}
\caption{Increase in pressure as fast wave approaches and crosses the  $c_s^2=v_A^2\:$ layer. The wave crosses the  $c_s^2=v_A^2\:$ layer at approximately $t=1.5$.}
\label{fig:2.3.3.5}
\end{center}
\end{figure}

The formation of the quasi-diamond shape in Figure  \ref{fig:2.3.3.2_MAGIC} is due to  a decrease in the overall wave  speed along the separatrices. This decrease is wave speed can be understood by investigating the  perturbed pressure, $p_1$, and this can be seen in Figure \ref{fig:2.3.3.4}. We see that $p_1$ propagates towards the null similar to the propagation of the fast wave and  is zero along the axes, i.e. the lines $x=0$ and $z=0$. Hence, because of the alternating nature of the pressure, the maximum gradients in pressure will occur along these locations, i.e. {\emph{along the separatrices}}. This pressure gradient  acts against the magnetic forces in the momentum equation and thus reduces the acceleration of the fast wave along the separatrices,  i.e. the magnitude of $\frac {\partial}{\partial t} {\rm{v}_\perp}$ is smaller along the separatrices  leading to the deceleration as seen in Figure \ref{fig:2.3.3.2_MAGIC}. Note also that the pressure is increasing all the time and this can be seen in Figure \ref{fig:2.3.3.5}.


Note that in this paper we do not describe the evolution of  ${\rm{v}_\perp}$ after it crosses the  $c_s^2=v_A^2\:$ layer; this crossing occurs at approximately $t=1.5$. As the wave crosses the  $c_s^2=v_A^2\:$ layer, complex MHD mode conversion occurs. However, the description of such mode conversion is not the focus of this current paper and  the resultant mode conversion has already been reported by McLaughlin \& Hood (\citeyear{MH2006b}). Instead, this paper focuses on (i) the nature of the wave propagation {\emph{before}} crossing the $c_s^2=v_A^2\:$ layer and (ii) comparing and contrasting this behaviour to that seen in the $\beta=0$ system. Thus, our main conclusion for the $\beta \neq 0$ system is related to the explanation of  the quasi-diamond shape in Figure  \ref{fig:2.3.3.2_MAGIC} and that this deformation in wave morphology was absent in the $\beta=0$ set-up. Note that McLaughlin \& Hood (\citeyear{MH2006b}) {\emph{does not}} include  our insights related to the formation of the quasi-diamond pattern as well as its explanation in terms of   the maximum gradients in pressure occurring along the separatrices. We  also note that early on in its evolution, the $\beta \neq 0$ fast wave evolves in a similar manner to its $\beta=0$ equivalent. By looking at the equations (\ref{finitebetaequations}), we see this makes sense; at large radii the pressure terms are negligible and so  the Alfv\'en speed is essentially spatially varying like  $r$, and so the refraction effect dominates the evolution.

We can also looked at the behaviour of ${\rm{v}_\parallel}$  in $\S\ref{ein2}$ and this can be seen in Figure \ref{fig:2.3.3.9}. We observe that the wave has an alternating structure in the $\theta$-direction, i.e.  positive and negative parts to the wave, unlike  ${\rm{v}_\perp}$ which was always positive, and we note that the ratio  ${\rm{v}_\parallel}$ : ${\rm{v}_\perp}$ of the amplitude of disturbances is $\beta_0$ : $1$. We also note that we have set an initial condition in ${\rm{v}}_\perp$ only: the initial condition on the parallel wave was  ${\rm{v}_\parallel}  (x,z,0) = 0$ in equations (\ref{initalconditionsIC}). Hence, the ${\rm{v}_\parallel}$ wave we are observing has  been generated as a consequence of our ${\rm{v}}_\perp$ initial condition. By looking at equations (\ref{finitebetaequations}) and our initial conditions, we see that   ${\rm{v}}_\perp$  acts as a driver or forcing term for ${\rm{v}_\parallel}$. Thus, we are  solving the equivalent of a second-order differential equation with a forcing term, which is an inhomogeneous  equation. The general solution to such equations consists of two parts; a {\emph{complementary function}} and a {\emph{particular integral}}. The  complementary function is a solution to the corresponding homogeneous differential equation whereas the particular integral is a solution to  the  inhomogeneous differential equation. Hence, returning our attention to  Figure \ref{fig:2.3.3.9}, we see that there should be two  parts to the ${\rm{v}_\parallel}$ wave. We do see a part which has the same speed and frequency as the perpendicular component wave and, using the definition above, this wave can be thought of as the particular integral to the equations. There is also be a complementary function part to the wave, though it is  difficult to see in the figure.

As a consequence  of our results from $\S\ref{sec:2.3.3}$ and $\S\ref{polarcoorindatesbetaneq0}$, we shall now explain how we  interprete the waves seen in the perpendicular and parallel velocities.


\subsection{Interpretating the waves we see in ${\bf{v}}_\perp$ and ${\bf{v}}_\parallel$}

Our MHD system contains three key velocities:  ${\bf{v}}_{\rm{Alfv{\acute{e}}n}}$,  ${\bf{v}}_{\rm{slow}}$ and ${\bf{v}}_{\rm{fast}}$, that are all orthogonal and thus we may consider them as an {\emph{orthogonal basis}} of vectors for our system. In this paper, we do not  consider the Alfv\'en wave, ${\bf{v}}_{\rm{Alfv{\acute{e}}n}}=v_y {\hat{\bf{y}}}$,  and so our 2D  vectors may be described in terms of the vectors   ${\bf{v}}_{\rm{fast}}$ and  ${\bf{v}}_{\rm{slow}}$. Due to our choice of coordinate system ($\S\ref{sec:1.2b}$) we choose to work in the directions perpendicular and parallel to the magnetic field. Thus, we may represent these two vectors in terms of ${\bf{v}}_{\rm{fast}}$ and  ${\bf{v}}_{\rm{slow}}$, namely
\begin{eqnarray*}
{\bf{v}}_\perp = {A} {\bf{v}}_{\rm{fast}} + {B}{\bf{v}}_{\rm{slow}}\;\; ,\quad {\bf{v}}_\parallel = {C} {\bf{v}}_{\rm{fast}} + {D}{\bf{v}}_{\rm{slow}}  \;.
\end{eqnarray*}
 Alternatively, we may express our two magnetoacoustic velocities in terms of ${\bf{v}}_\perp$ and ${\bf{v}}_\parallel$, namely
\begin{eqnarray*}
{\bf{v}}_{\rm{fast}} = E {\bf{v}}_\perp + F {\bf{v}}_\parallel \;\; ,\quad {\bf{v}}_{\rm{slow}} =G {\bf{v}}_\perp + H {\bf{v}}_\parallel  \;,
\end{eqnarray*}
where $A$, $B$, $C$, $D$, $E$, $F$, $G$ and $H$ are unknown functions that depend upon the magnetic geometry and (possibly)  the plasma-$\beta$. This representation is only possible because both ${\bf{v}}_{\rm{fast}}$ and ${\bf{v}}_{\rm{slow}}$ and  ${\bf{v}}_\perp$ and ${\bf{v}}_\parallel$  form  orthogonal bases.

However, we must be cautious: the concepts of fast and slow waves  were originally derived for a uni-directional magnetic field and so these ideas may not carry over to more complex magnetic  geometries quite as simply as claimed here. However, we recommend still utilizing terminology  such as fast and slow wave in the  interpretation of the waves in complex topologies, as well as the intuition gained from the uni-directional magnetic field models. We believe that a good way of interpreting the waves we see in our magnetic configuration  is as follows
\begin{eqnarray*}
\rm{fast \;wave  }&=&\rm{\;(large\; perpendicular \;component) \;}+\rm{\; (parallel \;component) }  \nonumber\\
\rm{ }&=&\rm{\;(large\; component\; in\; {{v}_\perp})\; }+\rm{\; (component \;in\; {{v}_\parallel}) }\; , \nonumber\\
\rm{slow \;wave\;  }&=&\rm{\; (small\; perpendicular \;component) \;}+\rm{ \;(parallel\; component)}  \nonumber\\
\rm{ }&=&\rm{\;(small\; component\; in\; {{v}_\perp})\; }+\rm{\;(component\; in \;{{v}_\parallel}) }\;.\label{interpretationslowfast}
\end{eqnarray*}

In addition, our system consists of a region of low-$\beta$ plasma outside the $\beta=1$ layer and a region of high-$\beta$ plasma within; see Figure \ref{fig:areasofhighandlow}. This is understood from our definition of the plasma-$\beta$ for this magnetic field; $\beta \propto \left( {x^2+z^2} \right) ^{-1}$. Recall that slow and fast waves have differing properties depending on whether they are in a high or low-$\beta$ environment. To summarise:
\vspace{0.25in}
\begin{table}[h]\label{FSprop}
\begin{center}
\begin{tabular}{|c|c|c|}
\hline
 & Fast Wave & Slow Wave\\
\hline\hline
High-$\beta$ & \begin{tabular}{c}
Behaves like sound wave\\
 (speed $c_s$)
\end{tabular} & 
\begin{tabular}{c}
Guided along ${\bf{B}}_0$ \\
 Transverse wave travelling at $v_A$
\end{tabular}\\
\hline
Low-$\beta$ & \begin{tabular}{c}
Propagates roughly isotropically \\
(speed $v_A$)
\end{tabular} & \begin{tabular}{c}
Guided along ${\bf{B}}_0$\\
 Longitudinal wave travelling at speed $c_T$\\
\end{tabular}\\
\hline
\end{tabular}
\end{center}
\end{table}

In our investigations, we have  sent a wave pulse into our system from a particular radius, i.e. in the low-$\beta$ region. At some point this wave has crossed the $\beta=1$ layer and entered the high-$\beta$ environment.  Thus, we  have a low-$\beta$ wave approaching the layer, coupling and mixing inside the layer and emerging as a mixture of high-$\beta$ fast and slow waves. We are driving waves in the perpendicular velocity component in a low-$\beta$ region (see Figure \ref{fig:areasofhighandlow}) and so we interpret this as predominantly low-$\beta$  fast wave.  At this time there does not exist a robust set of rules connecting low and high-$\beta$ waves across the $\beta=1$ layer. It is hoped that the work presented here will help contribute to such a set of rules, specifically in what happens when a low-$\beta$ fast wave crosses the $\beta=1$ layer and becomes part high-$\beta$ fast wave and part high-$\beta$ slow wave.

{{
We conclude that in a $\beta=0$ plasma, the fast wave cannot cross the null point and all the wave energy accumulates at that location. Thus, {\emph{null points will be locations for preferential heating from fast magnetoacoustic waves}}. For $\beta \neq 0$, the evolution is more complex and the fast wave now couples with the slow wave close to the $\beta=1$ layer. The resultant behaviour is controlled by the parameter $\beta_0$.

Finally, there is as yet no unambiguous observational evidence for MHD wave behaviour in the vicinity of coronal null points. The successful detection of MHD waves around coronal null points will require advancements in two areas: high-spatial and high-temporal resolution imaging data as well as magnetic extrapolations from co-temporal magnetograms. Future missions, such as the {\it{Daniel K. Inouye Solar Telescope}} and {\it{Solar Orbiter}} may satisfy these requirements, and so the first detection of MHD waves in the neighbourhood of null points may be reported in the near future.
}}


\bigskip
{\bf Acknowledgements}

{The author acknowledges IDL support provided by STFC. JM  wishes to thank Alan Hood for insightful discussions and constant encouragement.}



\end{document}